\documentclass[twocolumn]{aastex63}

\newcommand\tess{TESS}
\newcommand\gaia{\textit{Gaia}}
\newcommand\kms{$\textrm{km/s}$}
\newcommand\ms{$\textrm{m/s}$}
\newcommand\cms{$\textrm{cm/s}$}

\newcommand\teff{T$_{\rm{eff}}$}

\newcommand{\unit}[1]{\ensuremath{\, \mathrm{#1}}} 
\newcommand\earthmass{M$_{\oplus}$}
\newcommand\earthradius{R$_{\oplus}$}

\newcommand\plmass{$61.5_{-9.3}^{+9.7}$}
\newcommand\plradius{$5.82\pm0.19$}
\newcommand\stteff{$3957\pm69$}

\shortauthors{Kanodia et al. 2021.}
\shorttitle{A super Neptune orbiting TOI-532}

\received{June 16, 2021}
\accepted{July 28, 2021}

\begin{document}

\title{TOI-532b: The Habitable-zone Planet Finder confirms a Large Super Neptune in the Neptune Desert orbiting a metal-rich M dwarf host}

\author[0000-0001-8401-4300]{Shubham Kanodia}
\affiliation{Department of Astronomy \& Astrophysics, 525 Davey Laboratory, The Pennsylvania State University, University Park, PA, 16802, USA}
\affiliation{Center for Exoplanets and Habitable Worlds, 525 Davey Laboratory, The Pennsylvania State University, University Park, PA, 16802, USA}

\author[0000-0001-7409-5688]{Gudmundur Stefansson}
\affiliation{Henry Norris Russell Fellow}
\affiliation{Department of Astrophysical Sciences, Princeton University, 4 Ivy Lane, Princeton, NJ 08540, USA}

\author[0000-0003-4835-0619]{Caleb I. Ca\~nas}
\affiliation{NASA Earth and Space Science Fellow}
\affil{Department of Astronomy \& Astrophysics, 525 Davey Laboratory, The Pennsylvania State University, University Park, PA, 16802, USA}
\affil{Center for Exoplanets and Habitable Worlds, 525 Davey Laboratory, The Pennsylvania State University, University Park, PA, 16802, USA}

\author[0000-0001-8222-9586]{Marissa Maney}
\affil{Department of Astronomy \& Astrophysics, 525 Davey Laboratory, The Pennsylvania State University, University Park, PA, 16802, USA}
\affil{Center for Exoplanets and Habitable Worlds, 525 Davey Laboratory, The Pennsylvania State University, University Park, PA, 16802, USA}

\author[0000-0002-9082-6337]{Andrea S.J. Lin}
\affil{Department of Astronomy \& Astrophysics, 525 Davey Laboratory, The Pennsylvania State University, University Park, PA, 16802, USA}
\affil{Center for Exoplanets and Habitable Worlds, 525 Davey Laboratory, The Pennsylvania State University, University Park, PA, 16802, USA}

\author[0000-0001-8720-5612]{Joe P. Ninan}
\affil{Department of Astronomy \& Astrophysics, 525 Davey Laboratory, The Pennsylvania State University, University Park, PA, 16802, USA}
\affil{Center for Exoplanets and Habitable Worlds, 525 Davey Laboratory, The Pennsylvania State University, University Park, PA, 16802, USA}

\author[0000-0002-7227-2334]{Sinclaire Jones}
\affiliation{Department of Astrophysical Sciences, Princeton University, 4 Ivy Lane, Princeton, NJ 08540, USA}

\author[0000-0002-0048-2586]{Andrew Monson}
\affil{Department of Astronomy \& Astrophysics, 525 Davey Laboratory, The Pennsylvania State University, University Park, PA, 16802, USA}
\affil{Center for Exoplanets and Habitable Worlds, 525 Davey Laboratory, The Pennsylvania State University, University Park, PA, 16802, USA}

\author[0000-0001-9307-8170]{Brock A. Parker}
\affil{Department of Physics \& Astronomy, University of Wyoming, Laramie, WY 82070, USA}

\author[0000-0002-4475-4176]{Henry A. Kobulnicky}
\affil{Department of Physics \& Astronomy, University of Wyoming, Laramie, WY 82070, USA}

\author{Jason Rothenberg}
\affil{Department of Physics \& Astronomy, University of Wyoming, Laramie, WY 82070, USA}

\author[0000-0001-7708-2364]{Corey Beard}
\affil{Department of Physics and Astronomy, The University of California, Irvine, Irvine, CA 92697, USA}

\author[0000-0001-8342-7736]{Jack Lubin}
\affiliation{Department of Physics \& Astronomy, University of California Irvine, Irvine, CA 92697, USA}

\author[0000-0003-0149-9678]{Paul Robertson}
\affiliation{Department of Physics \& Astronomy, University of California Irvine, Irvine, CA 92697, USA}

\author[0000-0002-5463-9980]{Arvind F. Gupta}
\affil{Department of Astronomy \& Astrophysics, 525 Davey Laboratory, The Pennsylvania State University, University Park, PA, 16802, USA}
\affil{Center for Exoplanets and Habitable Worlds, 525 Davey Laboratory, The Pennsylvania State University, University Park, PA, 16802, USA}

\author[0000-0001-9596-7983]{Suvrath Mahadevan}
\affil{Department of Astronomy \& Astrophysics, 525 Davey Laboratory, The Pennsylvania State University, University Park, PA, 16802, USA}
\affil{Center for Exoplanets and Habitable Worlds, 525 Davey Laboratory, The Pennsylvania State University, University Park, PA, 16802, USA}

\author[0000-0001-9662-3496]{William D. Cochran}
\affiliation{McDonald Observatory and Department of Astronomy, The University of Texas at Austin}
\affiliation{Center for Planetary Systems Habitability, The University of Texas at Austin}

\author[0000-0003-4384-7220]{Chad F. Bender}
\affil{Steward Observatory, The University of Arizona, 933 N.\ Cherry Ave, Tucson, AZ 85721, USA}

\author[0000-0002-2144-0764]{Scott A. Diddams}
\affil{Time and Frequency Division, National Institute of Standards and Technology, 325 Broadway, Boulder, CO 80305, USA}
\affil{Department of Physics, University of Colorado, 2000 Colorado Avenue, Boulder, CO 80309, USA}

\author[0000-0002-0560-1433]{Connor Fredrick}
\affil{Time and Frequency Division, National Institute of Standards and Technology, 325 Broadway, Boulder, CO 80305, USA}
\affil{Department of Physics, University of Colorado, 2000 Colorado Avenue, Boulder, CO 80309, USA}

\author[0000-0003-1312-9391]{Samuel Halverson}
\affil{Jet Propulsion Laboratory, 4800 Oak Grove Drive, Pasadena, CA 91109, USA}

\author[0000-0002-6629-4182]{Suzanne Hawley}
\affil{Department of Astronomy, Box 351580, University of Washington, Seattle, WA 98195 USA}

\author[0000-0002-1664-3102]{Fred Hearty}
\affil{Department of Astronomy \& Astrophysics, 525 Davey Laboratory, The Pennsylvania State University, University Park, PA, 16802, USA}
\affil{Center for Exoplanets and Habitable Worlds, 525 Davey Laboratory, The Pennsylvania State University, University Park, PA, 16802, USA}

\author[0000-0003-1263-8637]{Leslie Hebb}
\affiliation{Department of Physics, Hobart and William Smith Colleges, 300 Pulteney Street, Geneva,
NY, 14456, USA}

\author[0000-0002-5893-2471]{Ravi Kopparapu}
\affiliation{NASA Goddard Space Flight Center, 8800 Greenbelt Road, Greenbelt, MD 20771, USA}
\affiliation{Sellers Exoplanet Environment Collaboration (SEEC), NASA Goddard Space Flight Center}

\author[0000-0001-5000-1018]{Andrew J. Metcalf}
\affiliation{Space Vehicles Directorate, Air Force Research Laboratory, 3550 Aberdeen Ave. SE, Kirtland AFB, NM 87117, USA}
\affiliation{Time and Frequency Division, National Institute of Technology, 325 Broadway, Boulder, CO 80305, USA} 
\affiliation{Department of Physics, 390 UCB, University of Colorado Boulder, Boulder, CO 80309, USA}

\author[0000-0002-4289-7958]{Lawrence W. Ramsey}
\affil{Department of Astronomy \& Astrophysics, 525 Davey Laboratory, The Pennsylvania State University, University Park, PA, 16802, USA}
\affil{Center for Exoplanets and Habitable Worlds, 525 Davey Laboratory, The Pennsylvania State University, University Park, PA, 16802, USA}

\author[0000-0001-8127-5775]{Arpita Roy}
\affiliation{Space Telescope Science Institute, 3700 San Martin Dr, Baltimore, MD 21218, USA}
\affiliation{Department of Physics and Astronomy, Johns Hopkins University, 3400 N Charles St, Baltimore, MD 21218, USA}

\author[0000-0002-4046-987X]{Christian Schwab}
\affil{Department of Physics and Astronomy, Macquarie University, Balaclava Road, North Ryde, NSW 2109, Australia}

\author[0000-0003-2435-130X]{Maria Schutte}
\affiliation{Homer L. Dodge Department of Physics and Astronomy, University of Oklahoma, 440 W. Brooks Street, Norman, OK 73019, USA}

\author[0000-0002-4788-8858]{Ryan C. Terrien}
\affil{Department of Physics and Astronomy, Carleton College, One North College Street, Northfield, MN 55057, USA}

\author[0000-0001-9209-1808]{John Wisniewski}
\affiliation{Homer L. Dodge Department of Physics and Astronomy, University of Oklahoma, 440 W. Brooks Street, Norman, OK 73019, USA}

\author[0000-0001-6160-5888]{Jason T.\ Wright}
\affiliation{Department of Astronomy \& Astrophysics, 525 Davey Laboratory, The Pennsylvania State University, University Park, PA, 16802, USA}
\affiliation{Center for Exoplanets and Habitable Worlds, 525 Davey Laboratory, The Pennsylvania State University, University Park, PA, 16802, USA}
\affiliation{Penn State Extraterrestrial Intelligence Center, 525 Davey Laboratory, The Pennsylvania State University, University Park, PA, 16802, USA}

\correspondingauthor{Shubham Kanodia}
\email{shbhuk@gmail.com}

\begin{abstract}
We confirm the planetary nature of TOI-532b, using a combination of precise near-infrared radial velocities with the Habitable-zone Planet Finder, TESS light curves, ground based photometric follow-up, and high-contrast imaging. TOI-532 is a faint (J$\sim 11.5$) metal-rich M dwarf with \teff{} = \stteff{} K and [Fe/H] = $0.38\pm0.04$; it hosts a transiting gaseous planet with a period of $\sim 2.3$ days. Joint fitting of the radial velocities with the \tess{} and ground-based transits reveal a planet with radius of \plradius{} \earthradius{}, and a mass of \plmass{} \earthmass{}. TOI-532b is the largest and most massive super Neptune detected around an M dwarf with both mass and radius measurements, and it bridges the gap between the Neptune-sized planets and the heavier Jovian planets known to orbit M dwarfs. It also follows the previously noted trend between gas giants and host star metallicity for M dwarf planets. In addition, it is situated at the edge of the Neptune desert in the Radius--Insolation plane, helping place constraints on the mechanisms responsible for sculpting this region of planetary parameter space. 

\end{abstract}

\keywords{planets and satellites: detection, composition; planetary systems; stars: fundamental parameters; methods: statistical;}

\section{Introduction} \label{sec:intro}

Studies analyzing the host star metallicity dependence of gas giant ($R_p > 4$ \earthradius{}) occurrence rates have traditionally relied on a sample of planets orbiting solar type stars, with a typical minimum photospheric temperature corresponding to mid K dwarfs. Extending this analysis to M dwarf planets has been hampered by the intrinsic faintness of M dwarfs, which makes planet detection and mass measurement difficult. Occurrence rate studies for transiting planets orbiting M dwarfs have been limited to the smaller ($R_p < 4$ \earthradius{}) planets \citep{laughlin_core_2004, johnson_metal_2009, gaidos_understanding_2013, dressing_occurrence_2015, hsu_occurrence_2020}. Attempts to study the occurrence rates of gas giants orbiting M dwarfs have used samples from radial velocity (RV) surveys \citep{johnson_metal_2009, johnson_giant_2010, gaidos_understanding_2013, tuomi_frequency_2019}. Most recently, \cite{maldonado_hades_2020} use a sample of RV planets detected from the HARPS-N spectrograph to probe the dependence of gas giant occurrence on metallicity. Occurrence rate studies for gaseous planets using RV surveys can be complicated by the lack of true mass measurements ($M_p$ vs $M_p$ sin\textit{i}).  Therefore, in its all-sky survey of transiting planets around nearby-stars---and with its red-optimized band-pass yielding high precision photometric observations of nearby M-dwarfs--- the \textit{Transiting Exoplanet Survey Satellite}  \citep[\tess{};][]{ricker_transiting_2014} presents a unique opportunity to find transiting gas giants orbiting M dwarfs suitable for mass measurements. Four such recent discoveries by TESS are---TOI-1728b \citep{kanodia_toi-1728b_2020}, TOI-1899b \citep{canas_warm_2020}, TOI-442b \citep{dreizler_carmenes_2020}, and TOI-674b \citep{murgas_toi-674b_2021}.

Transiting Neptune-sized planets ($2 R_{\oplus} < R_p < 6 R_{\oplus}$)\footnote{Also referred to as sub-Saturns \citep{petigura_california-keplersurvey_2018, kopparapu_exoplanet_2018}.}, present a transitional population between rocky terrestrial planets and Jovian gas giants. In particular, transiting super Neptunes \citep[$17 M_{\oplus} < M_p < 57 M_{\oplus}$;][]{bakos_hats-7b_2015}, can help inform theories of planet formation and migration, i.e., did the gaseous giants form in-situ close to their host star, or form away beyond the ice line and migrate inwards due to eccentricity driven excitation or disk migration \citep{madhusudhan_atmospheric_2017, bean_nature_2021, fortney_hot_2021}. This investigation into the provenance of gaseous giants can be further aided by atmospheric characterization using transmission spectroscopy \citep{guzman-mesa_information_2020}, where the ``warm Neptunes" with equilibrium temperatures between $\sim 800$-1200 K, are expected to exhibit diverse atmospheric elemental abundances, with possible imprints of the protoplanetary disk chemistry \citep{mordasini_imprint_2016}. 

Additionally, as predicted by \cite{ida_toward_2004}, \cite{szabo_short-period_2011} and \cite{mazeh_dearth_2016} have noted a dearth of Neptune-sized objects orbiting close to their host star (2-4 day orbital period), referred to as the ``Neptune Desert". Different hypotheses have been proposed as a possible explanation to this feature, since it can not be explained by observational biases. \cite{matsakos_origin_2016} attempt to explain the origin of the Neptune Desert using high eccentricity migration, whereas \cite{owen_photoevaporation_2018} show that photoevaporation can be a driving factor responsible for the lower boundary of the desert. 

In this manuscript, we report the discovery of the transiting Super Neptune TOI-532b using precision RVs from the near infrared (NIR) Habitable-zone Planet Finder spectrograph \citep[HPF;][]{mahadevan_habitable-zone_2012, mahadevan_habitable-zone_2014}, to measure the mass of a transiting super Neptune orbiting the early type metal-rich M dwarf TOI-532 in the constellation of Orion. We perform a comprehensive characterization of the stellar and planetary properties using space-based photometric observations from TESS, additional ground-based transit observations, adaptive optics imaging, and high-contrast speckle imaging. This paper is structured as follows. In Section \ref{sec:observations}, we discuss the observations of this system, which include space-based TESS photometry, ground-based photometry, high contrast imaging, as well as precision RV observations with HPF. In Section \ref{sec:stellar} we discuss our characterization of the stellar parameters, followed by Section \ref{sec:joint}, where we detail our joint analysis of the photometry and velocimetry to constrain the planetary parameters of TOI-532b. In Section \ref{sec:discussion}, we compare the properties of TOI-532b with other M dwarf exoplanets, and with few other Neptunes to place it in context for potential He 10830 \AA~absorption detection using transmission spectroscopy. Finally, we summarize our results in Section \ref{sec:summary}. 

\begin{figure*}[] 
\centering
\includegraphics[width=\textwidth]
{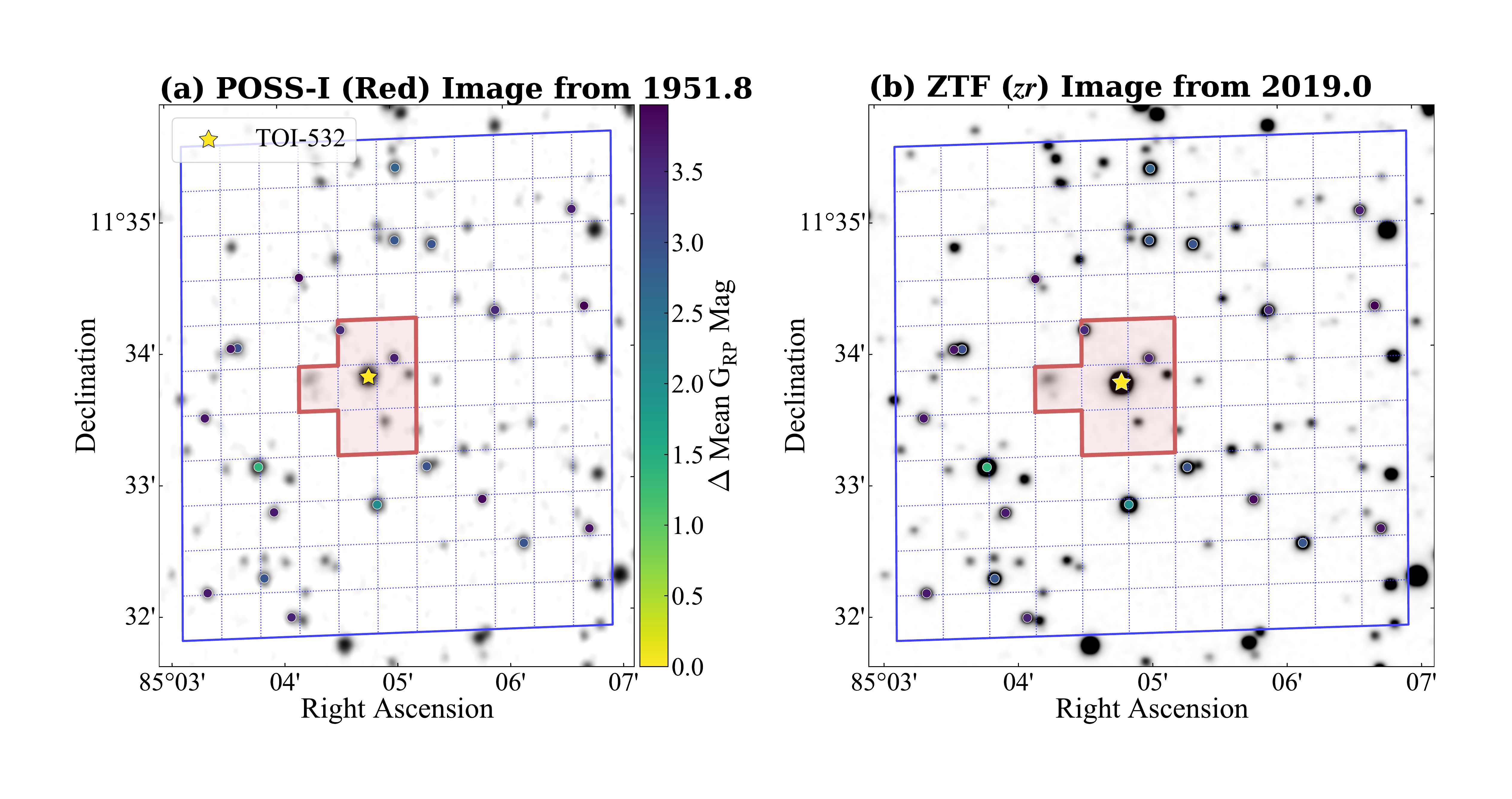}
\caption{\textbf{Panel a} overlays the 11 x 11 pixel \tess{} Sector 6 footprint (blue grid) on a POSS-I red image from 1951.8. TOI-532 does not have significant proper motion, as can be seen while comparing Panel \textbf{a)} and \textbf{b)}. The \tess{} aperture is outlined in red and we highlight our target TOI-532. No bright targets are present inside the \tess{} aperture with $\Delta$ G$_{RP}$ $<$ 3. \textbf{Panel b} is similar to Panel A but with a background image from ZTF \(zr\) (5600 \AA -- 7316 \AA) 2019 \citep{masci_zwicky_2019}.} \label{fig:tess_map}
\end{figure*}

\section{Observations}\label{sec:observations}
\subsection{TESS}\label{sec:TESS}
\begin{figure*}[!t] 
\centering
\includegraphics[width=1\textwidth]{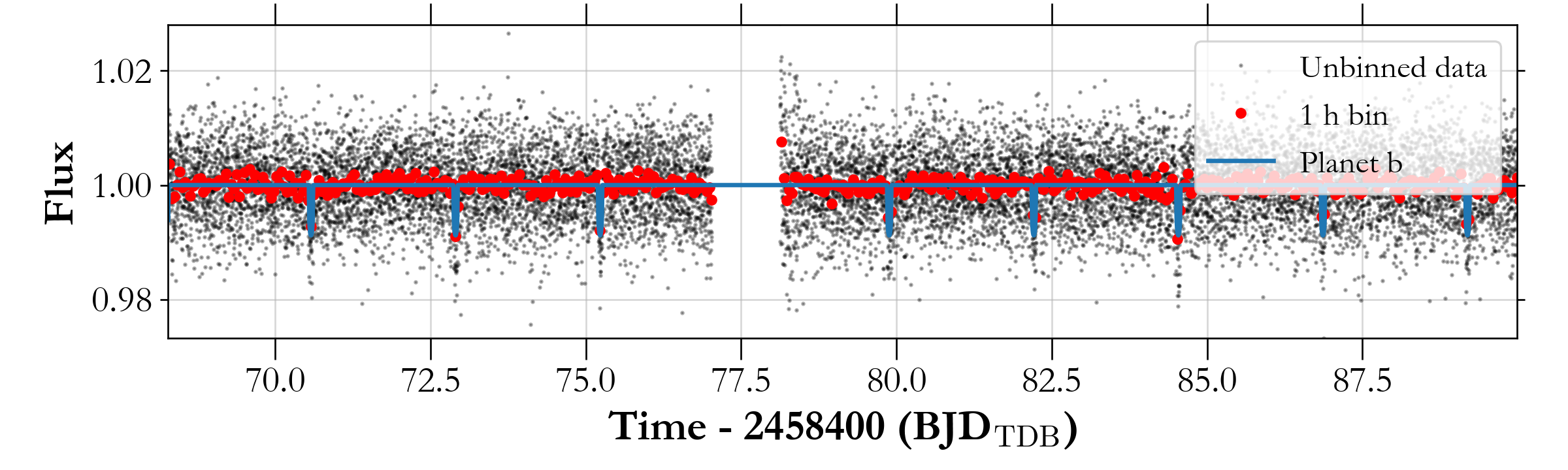}
\caption{Short cadence (2 minute) time series \tess{} PDCSAP photometry (without detrending) from Sector 6, with the binned data (in 1 hour   bins), along with the TOI-532b transits overlaid in blue. \explain{Updated X axis label}} \label{fig:tess_lc}
\end{figure*}

TOI-532 (TIC-144700903, 2MASS J05401918+1133463, \gaia{} EDR3 3340265717587057536, UCAC4 508-014156) was observed by TESS in Sector 6 in Camera 1 from 2018 December 11 to 2019 January 7th at two minute cadence (Figure \ref{fig:tess_lc}). The Science Processing Operations Center (SPOC) at NASA Ames \citep{jenkins_tess_2016} reported one transiting planet candidate, TOI-532.01, with a period of 2.326811 days. For our subsequent analysis, we use the Presearch Data Conditioning Single Aperture Photometry (PDCSAP) lightcurve, which contains systematics and dilution corrected data using the algorithms originally developed for the \textit{Kepler} data analysis pipeline. We retrieved the data using the  \texttt{lightkurve} package \citep{lightkurve_collaboration_lightkurve_2018}, available at the Mikulski Archive for Space Telescopes (MAST).

\autoref{fig:tess_map}  presents a comparison of the region contained within the Sector 6 footprint from the Palomar Observatory Sky Survey \citep[POSS-1;][]{harrington_48-inch_1952, minkowski_national_1963} image in 1951 and a more recent ZTF \citep{masci_zwicky_2019} image from 2019. There are no bright targets with $\Delta$ G$_{RP}$ $<$ 3 present in the \tess{} aperture, however there are a few targets with $\Delta$ G$_{RP}$ $<$ 4, that dilute the \tess{} transit. Even though this is taken into account in the PDCSAP flux, following the methodology of \cite{burt_toi-824_2020} we use our ground based photometry to estimate an additional correction to this dilution term photometry and discuss this in Section \ref{sec:joint}.

\subsection{Ground Based Photometric Follow up}\label{sec:photometry}

\begin{figure*}[!t] 
\centering
\includegraphics[width=1.1\textwidth]{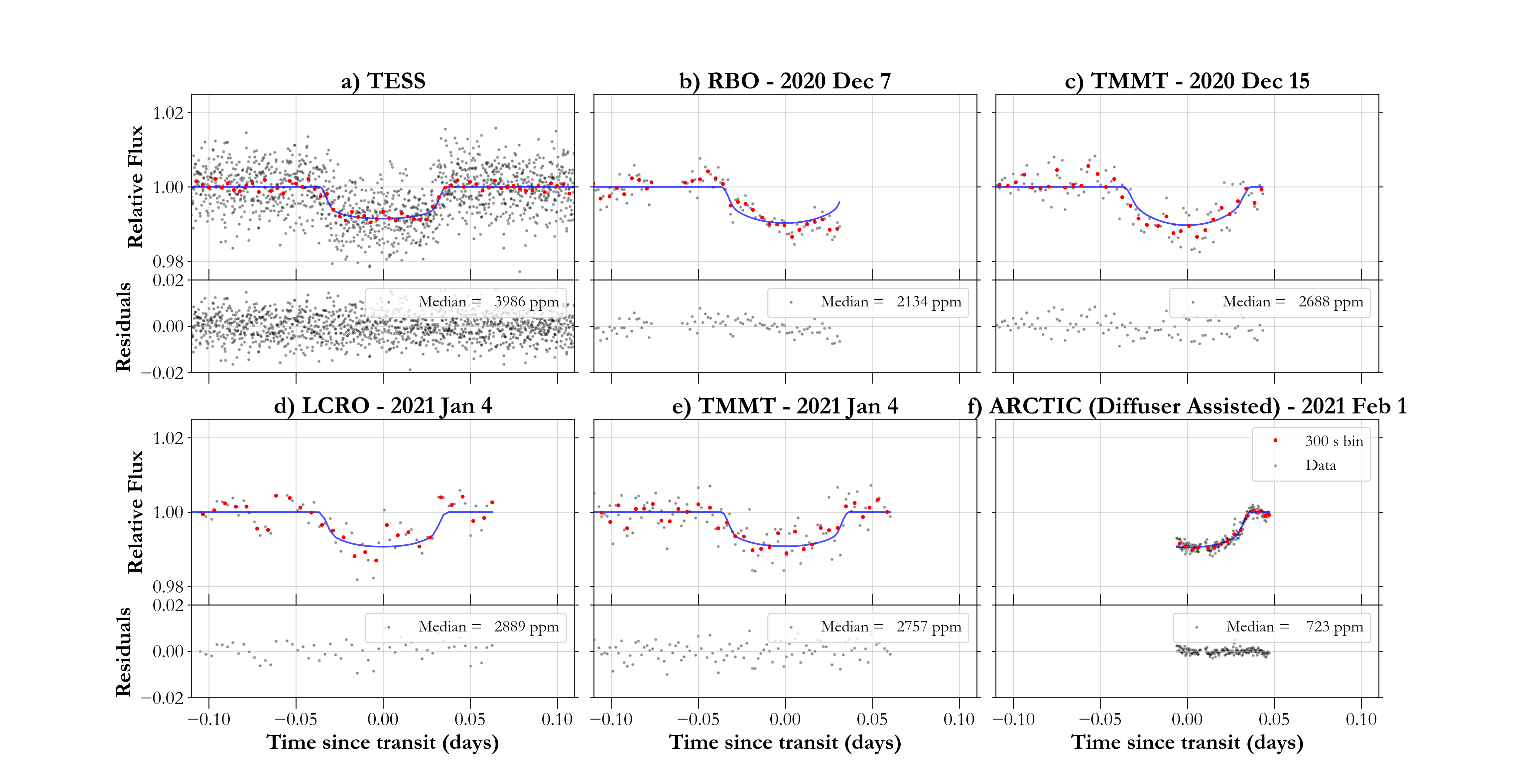}
\caption{Photometric observations for TOI-532b; a) the TESS phased plot shows the light curve phase-folded to the best fit orbital period, b-f) Ground based observations for TOI-532b. The raw photometery is shown in grey, whereas in red we show the photometry binned to 5 minute bins. The best-fit transit solution, along with the 1 $\sigma$ confidence interval are shown in blue.} \label{fig:transits}
\end{figure*}

We obtain follow up transits from the ground, to validate the transit seen in the \tess{} photometry, and measure the dilution present therein. Furthermore, the ground based photometry helps in improving the radius estimates as well as the ephemeris. These observations were pipeline processed using standard linearity, bias, dark, and flat field corrections. We then performed aperture photometry using AstroImageJ \citep{collins_astroimagej_2017}. Clear outliers due to cosmic rays, charged-particle events, poor seeing conditions, or telescope tracking were removed using AstroImageJ. We experimented using a number of different aperture settings, and varied the radii of the photometric aperture, as well as the inner, and outer background annuli, and selected the settings that resulted in the minimum scatter in the resulting photometry. Following the methodology in \cite{stefansson_toward_2017}, we added the scintillation error estimates to the photometric error estimated by AstroImageJ. See \autoref{fig:transits}, and \autoref{tab:photometric} for a summary of all our ground based photometric observations.
 
\subsubsection{RBO}
We observed a transit of TOI-532b on the night of 2020 December 7 using the $0.6 \unit{m}$ telescope at the Red Buttes Observatory (RBO) in Wyoming \citep{kasper_remote_2016}. The RBO telescope is a f/8.43 Ritchey-Chrétien Cassegrain constructed by DFM Engineering, Inc.  It is currently equipped with an Apogee ASPEN CG47 camera.

The target rose from an airmass of 1.61 at the start of the observations to a minimum airmass of 1.15 and then set to an airmass of 1.20 at the end of the observations. Observations were performed using the Bessell I filter \citep{bessell_ubvri_1990} with $1 \times 1$ on-chip binning. To prevent saturation, we defocused moderately (\autoref{tab:photometric}), which allowed us to use an exposure time of $120 \unit{s}$. In the $1 \times 1$ binning mode, the $0.6 \unit{m}$ at RBO has a gain of $1.27 \unit{e/ADU}$, a plate scale of $0.532 \arcsec$, and a readout time of approximately $2.4 \unit{s}$. 

 Due to cloud contamination, only the transit ingress was recovered from these observations (Figure \ref{fig:transits}b). For the final reduction, we selected a photometric aperture of 17 pixels (9.04$\arcsec$) with an inner sky annulus of 40 pixels (21.3$\arcsec$) and outer sky annulus of 60 pixels (31.9$\arcsec$).

\subsubsection{TMMT}
We observed two transits of TOI-532b  on the nights of 2020 December 15 (Figure \ref{fig:transits}c) and 2021 January 4 (Figure \ref{fig:transits}e) using the Three-hundred MilliMeter ($300 \unit{mm}$) Telescope \citep[TMMT;][]{monson_standard_2017} at  Las Campanas Observatory in Chile. TMMT is a f/7.8 FRC300 from Takahashi on a German equatorial AP1600 GTO mount with an Apogee Alta U42-D09 CCD Camera,  FLI ATLAS focuser, and Centerline filter wheel.

On 2020 December 15, the target rose from an airmass of 1.86 at the start of the observations to a minimum airmass of 1.32 and then set to an airmass of 2.62 at the end of the observations. On 2021 January 4, the target rose from an airmass of 1.48 to a minimum airmass of 1.32 and then set to an airmass of 3.16 at the end of observations. Observations on both nights were performed using the Bessell I filter with $1 \times 1$ on-chip binning and exposure times of 120 $\unit{s}$. In the $1 \times 1$ binning mode, TMMT has a gain of $1.35 \unit{e/ADU}$, a plate scale of $1.194 \unit{\arcsec/pixel}$, and a readout time of $6 \unit{s}$.

In addition to the standard corrections, a fringe subtraction was also performed for the TMMT I band images. The final light curve from 2020 December 15 utilized a photometric aperture of 5 pixels (5.97$\arcsec$), an inner sky annulus of 20 pixels (23.9$\arcsec$), and a outer sky annulus of 30 pixels (35.8$\arcsec$). The final light curve from 2021 January 4 utilized a photometric aperture of 5 pixels (5.97$\arcsec$), an inner sky annulus of 15 pixels (17.9$\arcsec$) and outer sky annulus of 30 pixels (35.8$\arcsec$). 

\subsubsection{LCRO}
We observed a transit of TOI-532b on the night of 2021 January 4 (Figure \ref{fig:transits}d) using the $305 \unit{mm}$ Las Campanas Remote Observatory (LCRO) telescope at the Las Campanas Observatory in Chile. The LCRO telescope is an f/8 Maksutov-Cassegrain from Astro-Physics on a German Equatorial AP1600 GTO mount with an FLI Proline 16803 CCD Camera, FLI ATLAS focuser and Centerline filter wheel.

The target rose from an airmass of 1.40 at the start of the observations to a minimum airmass of 1.32 and then set to an airmass of 3.29 at the end of the observations. Observations were performed using the SDSS $i^\prime$ filter with $1 \times 1$ on-chip binning and an exposure time of $240 \unit{s}$. In the $1 \times 1$ binning mode, LCRO has a gain of $1.52 \unit{e/ADU}$, and a plate scale of $0.773 \unit{\arcsec/pixel}$, and a readout time of $17 \unit{s}$. For the final reduction, we selected a photometric aperture of 6 pixels (4.64$\arcsec$) with an inner sky annulus of 13 pixels (10.0$\arcsec$) and outer sky annulus of 30 pixels (23.2$\arcsec$).

\subsubsection{Diffuser-assisted Photometry with the 3.5m ARC Telescope}
We observed a transit of TOI-532b (Figure \ref{fig:transits}f) on the night of 2021 February 1 using the $3.5\unit{m}$ Astrophysical Research Consortium (ARC) Telescope Imaging Camera \citep[ARCTIC;][]{huehnerhoff_astrophysical_2016} at the ARC 3.5m Telescope at Apache Point Observatory (APO). We observed the transit using the Engineered Diffuser available on ARCTIC, which we designed to enable precision photometric observations from the ground on nearby bright stars \citep{stefansson_toward_2017}.

The target set from an airmass of 1.07 at the start of the observations to an airmass of 1.14 at the end of the observations. The observations were performed using the SDSS $i^\prime$ filter with an exposure time of 20 $\unit{s}$ in the LL-readout and fast readout modes with $4 \times 4$ on-chip binning. In the $4 \times 4$ binning mode, ARCTIC has a gain of $2.0 \unit{e/ADU}$, a plate scale of $0.456 \unit{\arcsec/pixel}$, and a readout time of  $2.7 \unit{s}$. Due to cloud contamination, only the egress of the transit was recovered from the data. For the final reduction, we selected a photometric aperture of 13 pixels (5.72$\arcsec$), an inner sky annulus of 30 pixels (13.2$\arcsec$), and outer sky annulus of 45 pixels (19.8$\arcsec$).

\begin{deluxetable*}{cccccc}
\tablecaption{Summary of ground based photometric follow up of TOI-532 \label{tab:photometric}}
\centering
\tablehead{\colhead{Obs Date} & \colhead{Filter} & \colhead{Exposure} & \colhead{PSF} & \colhead{Apertures: Photometric,} & \colhead{Field of View} \\
\colhead{(YYYY-MM-DD)} & \colhead{} & \colhead{Time (s)} & \colhead{FWHM (")} & \colhead{Inner, Outer Annuli (")} & \colhead{(')} } 
\startdata
\multicolumn{6}{c}{\hspace{-0.2cm}RBO (0.6 m)}  \\
2020-12-07 & Bessell I & 120 & 8.88 (Defocus) & 9.04, 21.3, 31.9 & 8.94 $\times$ 8.94 \\
\hline
\multicolumn{6}{c}{\hspace{-0.2cm}TMMT (0.3 m)}  \\
2020-12-15 & Bessell I & 120 & 3.49 & 5.97, 23.9, 35.8 & 40.75 $\times$ 40.75 \\
2021-01-04 & Bessell I & 120 & 3.18 & 5.97, 17.9, 35.8 & 40.75 $\times$ 40.75 \\
\hline
\multicolumn{6}{c}{\hspace{-0.2cm}LCRO (0.3 m)}  \\
2021-01-04 & \textit{i'} & 240 & 2.45 & 4.64, 10.0, 23.2 & 51.97 $\times$ 51.97 \\
\hline
\multicolumn{6}{c}{\hspace{-0.2cm}APO (3.5 m)}  \\
2021-02-01 & \textit{i'} & 20 & 7.67 (Diffuser$^a$) & 5.72, 13.2, 19.8 &  7.9 $\times$ 7.9 \\
\enddata
\tablenotetext{a}{Engineered diffuser with 8.7$\arcsec$ FWHM \citep{stefansson_toward_2017}}
\end{deluxetable*}

\subsection{High Contrast Imaging}

\begin{figure}[] 
\centering
\includegraphics[width=0.5\textwidth]{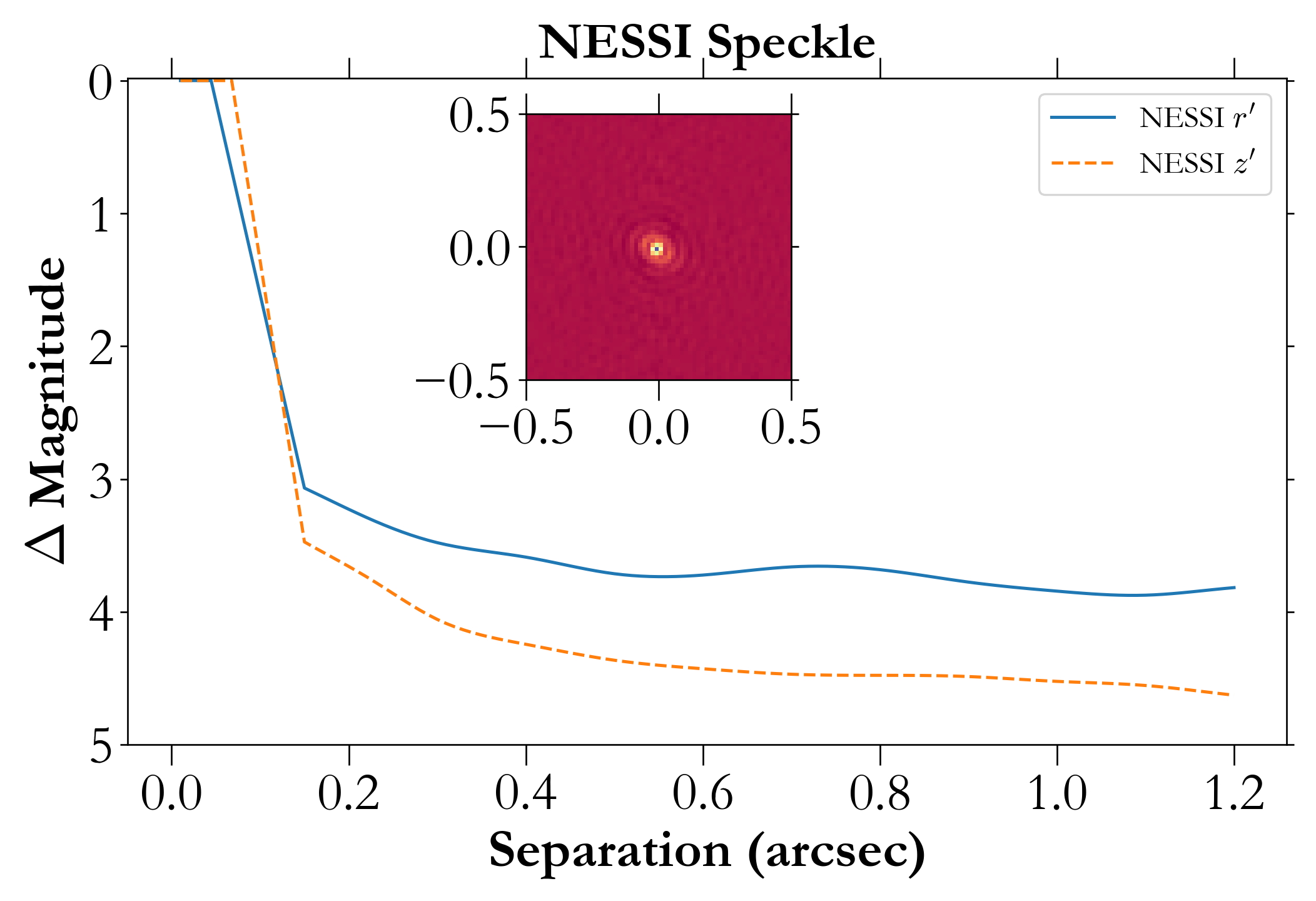}
\caption{5$\sigma$ contrast curve for TOI-532 observed from NESSI in the Sloan \(r^\prime\) and \(z^\prime\) filters showing no bright companions within \(1.2''\) from the host star. The \(z^\prime\) image is shown as an inset 1$\arcsec$ across.} \label{fig:NESSI}
\end{figure}

\begin{figure}[] 
\centering
\includegraphics[width=0.46\textwidth]{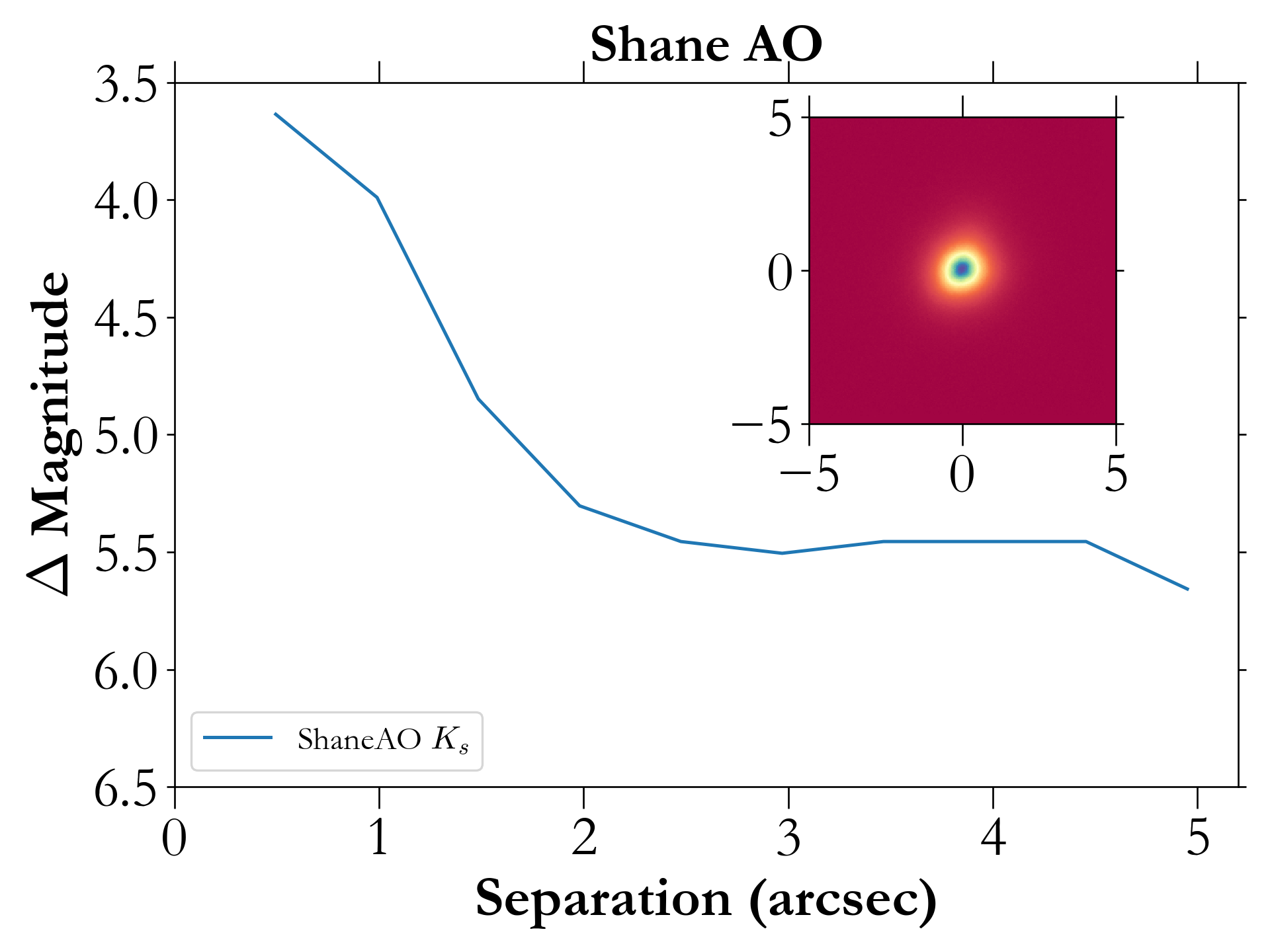}
\caption{5$\sigma$ contrast curve for TOI-532 from the ShARCS camera on the Shane 3 m telescope. We detected no companions within 0.507 $\pm$ 0.017 $\arcsec$ corresponding to a $\Delta$Ks of 3.7. The inset shows a 10 $\arcsec$ region around the star.} \label{fig:ShaneAO}
\end{figure}

\subsubsection{ShARCS on the Shane telescope}\label{sec:shane}

We observed TOI-532 using the ShARCS camera on the Shane 3m telescope at Lick Observatory \citep{srinath_swimming_2014}. Due to instrument repairs, we were unable to use the Laser Guide Star (LGS) mode, and had to use Natural Guide Star (NGS) mode. This mode can be more challenging for faint targets, as the guider camera can easily lose the target, but conditions were good enough to retrieve data for TOI-532. The target was observed using a 5 point dither process as outlined in \cite{furlan_kepler_2017}.

The data is then reduced using a custom AO pipeline developed internally. This pipeline first rejects all overexposed or underexposed images, and we then manually exclude data we know to be erroneous (lost guiding on the star, shutters closed early due to weather, etc.). Next we apply a standard dark correction, flat correction, and sigma clipping process. A master sky image is produced from the 5 point dither process, and subtracted from each image. A final image is then produced using an interpolation process to shift the images onto a single centroid.

Finally, we use the algorithm developed by \cite{espinoza_hats-25b_2016} to generate a 5 sigma contrast curve as a part of the final analysis (\autoref{fig:ShaneAO}). We detected no companions within $>$ 0.507 arcseconds corresponding to a $\Delta$Ks of 3.7.

\subsubsection{NESSI at WIYN}\label{sec:nessi}

We supplement our AO data with speckle imaging observations taken on 2021 April 3 using the NN-Explore Exoplanet Stellar Speckle Imager (NESSI) on the WIYN 3.5m telescope at Kitt Peak National Observatory. To search for faint background stars and stellar \replaced{comapnions}{companions}, we collected a 9 minute sequence of 40 ms diffraction-limited exposures of TOI-532 with the Sloan \(r^\prime\) and \(z^\prime\) filters. As we show in \autoref{fig:NESSI}, the NESSI data show no evidence of blending from a bright companion at separations $> 0.15$\arcsec at $\Delta r^{\prime}$ = 3.1, and $\Delta z^{\prime}$ = 3.5.

\begin{figure*}[!t] 
\centering
\includegraphics[width=\textwidth]{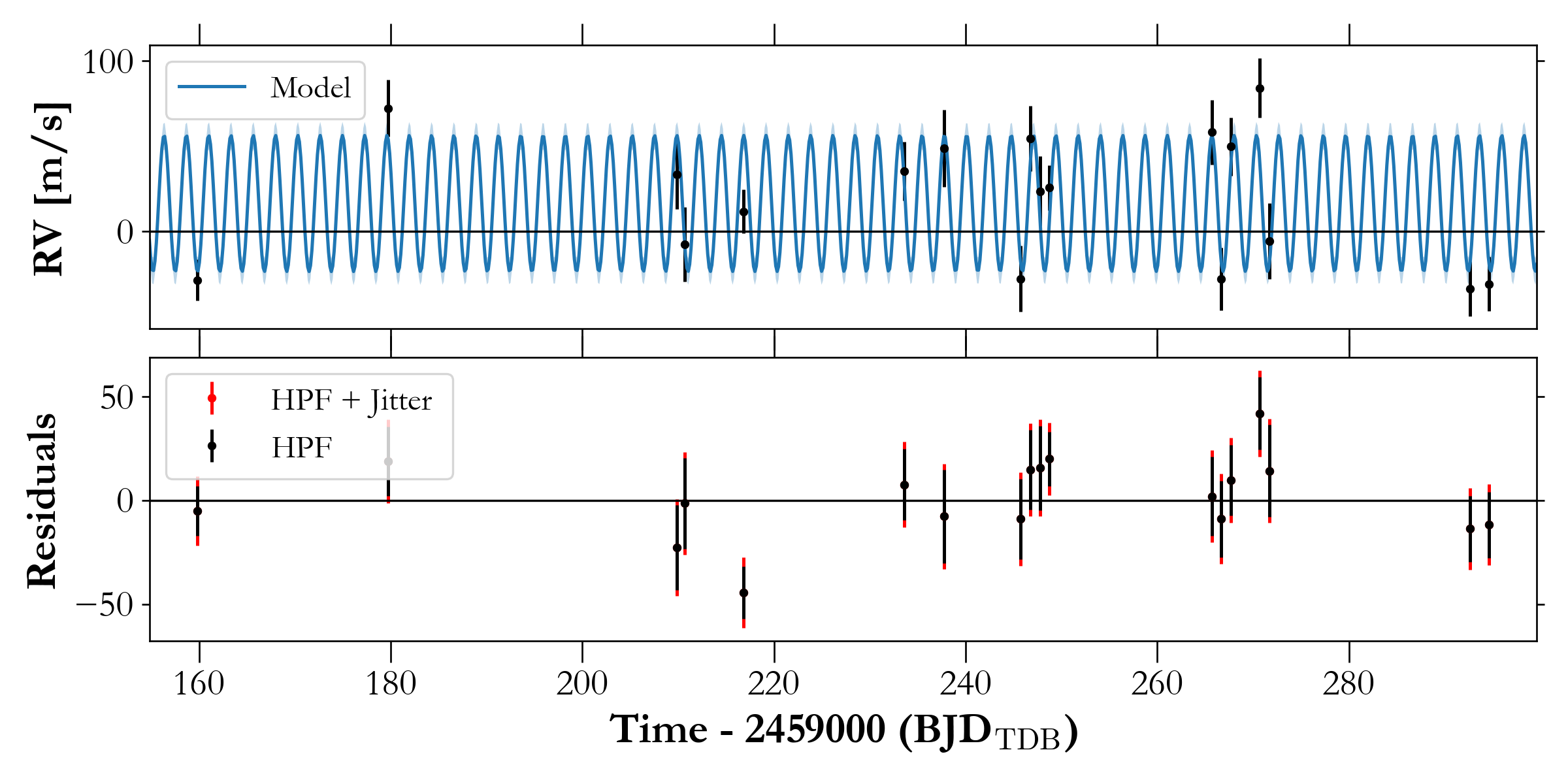}
\caption{Time series of RV observations of TOI-532 with HPF. The best-fitting model derived from the joint fit to the photometry and RVs is plotted in blue, including the 16-84$\%$ confidence interval in light blue. The bottom panel shows the residuals after subtracting the model. The jitter that is added in quadrature to the HPF errorbars (is shown in red), and is negligible compared to the HPF errorbars. \explain{Updated X axis label}} \label{fig:rv}
\end{figure*}

\subsection{Radial Velocity Follow-up with the Habitable-zone Planet Finder}\label{sec:hpfrvs}

We observed TOI-532 using HPF \citep{mahadevan_habitable-zone_2012, mahadevan_habitable-zone_2014}, a near-infrared (\(8080-12780\)\ \AA), high resolution precision RV spectrograph located at the 10 meter Hobby-Eberly Telescope (HET) in Texas. HET is a fixed-altitude telescope with a roving pupil design. It is fully queue-scheduled telescope with all observations executed in a queue by the HET resident astronomers \citep{shetrone_ten_2007}. HPF is a fiber-fed instrument with a separate science, sky and simultaneous calibration fiber \citep{kanodia_overview_2018}, and is actively temperature-stabilized at the milli-Kelvin level \citep{stefansson_versatile_2016}. We use the algorithms described in the tool \texttt{HxRGproc} for bias removal, non-linearity correction, cosmic ray correction, slope/flux and variance image calculation \citep{ninan_habitable-zone_2018} of the raw HPF data. HPF has the capability for simultaneous calibration using a NIR Laser Frequency Comb \citep[LFC;][]{metcalf_stellar_2019}, however owing to the faintness of our target we chose to avoid simultaneous calibration to minimize the impact of scattered calibrator light in the science target spectra. Instead, we interpolate the wavelength solution from other LFC exposures on the night of the observations, to correct for the well calibrated instrument drift, as has been discussed in \cite{stefansson_sub-neptune-sized_2020}. This method has been shown to enable precise wavelength calibration and drift correction with a precision of $\sim30$ \cms{} per observation, a value much smaller than our estimated per observation RV uncertainty (instrumental + photon noise) for this object of $\sim 22$ \ms{} (in 649 s exposures). 

To estimate the RVs, we follow the method described in \cite{stefansson_sub-neptune-sized_2020}, by using a modified version  of the \texttt{SpEctrum Radial Velocity AnaLyser} pipeline \citep[\texttt{SERVAL};][]{zechmeister_spectrum_2018}. \texttt{SERVAL} uses the template-matching technique to derive RVs \citep[e.g.,][]{anglada-escude_harps-terra_2012}, where it creates a master template from the target star observations, and determines the Doppler shift for each individual observation by minimizing the \(\chi^2\) statistic. We create this master template by using all the HPF observations of TOI-532, where telluric and sky-emission lines are masked in the calculations of the RVs. The telluric regions identified by a synthetic telluric-line mask generated from \texttt{telfit} \citep{gullikson_correcting_2014}, a Python wrapper to the Line-by-Line Radiative Transfer Model package \citep{clough_atmospheric_2005}.  Given the faintness of our target, we do not subtract out the sky fiber spectra from the sky fiber, as we observed that doing so added additional read noise, resulting in less precise RV measurements. To perform our barycentric correction, we use \texttt{barycorrpy}, the Python implementation \citep{kanodia_python_2018} of the algorithms from \cite{wright_barycentric_2014}. We obtained a total of 19 visits on this target between 2020 November 5 and 2021 March 21,  of which 1 visit was discarded due to bad weather conditions. Each visit was divided into 3 exposures of 649 seconds each, where the median S/N of each HPF exposure was 37 per resolution element. The individual exposures were then binned after weighting, with the final binned RVs being listed in \deleted{Table} \autoref{tab:rvs} and plotted in \autoref{fig:rv}.

\begin{deluxetable}{cccc}
\tablecaption{HPF RVs of TOI-532.  We include this table in a machine readable format along with the manuscript.\label{tab:rvs}}
\tablehead{\colhead{$\unit{BJD_{TDB}}$}  &  \colhead{RV}   & \colhead{$\sigma$} \\
           \colhead{(d)}   &  \colhead{$(\unit{m/s})$} & \colhead{$(\unit{m/s})$} }
\startdata
2459159.82217 & -28.41 & 12.02 \\ 
2459179.76468 & 72.25 & 16.71 \\ 
2459209.85498 & 33.54 & 20.41 \\ 
2459210.67441 & -7.51 & 21.93 \\ 
2459216.82807 & 11.79 & 12.70 \\ 
2459233.61084 & 35.41 & 17.17 \\ 
2459237.77469 & 48.57 & 22.57 \\ 
2459245.75139 & -27.58 & 19.35 \\ 
2459246.74596 & 54.53 & 19.16 \\ 
2459247.75757 & 23.59 & 20.39 \\ 
2459248.74084 & 25.75 & 13.16 \\ 
2459265.69221 & 58.13 & 19.00 \\ 
2459266.69484 & -27.78 & 18.47 \\ 
2459267.68911 & 49.75 & 16.90 \\ 
2459270.68444 & 84.04 & 17.46 \\ 
2459271.68015 & -5.66 & 22.22 \\ 
2459292.62161 & -33.47 & 15.94 \\ 
2459294.61636 & -30.69 & 15.83 \\
\enddata
\end{deluxetable}

\section{Stellar Parameters}\label{sec:stellar}

\begin{deluxetable*}{lccc}
\tablecaption{Summary of stellar parameters for TOI-532. \label{tab:stellarparam}}
\tablehead{\colhead{~~~Parameter}&  \colhead{Description}&
\colhead{Value}&
\colhead{Reference}}
\startdata
\multicolumn{4}{l}{\hspace{-0.2cm} \textit{Main identifiers:}}  \\
~~~TOI & \tess{} Object of Interest & 532 & \tess{} mission \\
~~~TIC & \tess{} Input Catalogue  & 144700903 & Stassun \\
~~~2MASS & \(\cdots\) & J05401918+1133463 & 2MASS  \\
~~~WISE &\(\cdots\) & J054019.20+113345.6 & WISE \\
~~~\gaia{} EDR3 & \(\cdots\) & 3340265717587057536 & \gaia{} EDR3\\
\multicolumn{4}{l}{\hspace{-0.2cm} \textit{Equatorial Coordinates, Proper Motion and Spectral Type:}} \\
~~~$\alpha_{\mathrm{J2016}}$ &  Right Ascension (RA, degrees) & 85.08005702(4) & \gaia{} EDR3\\
~~~$\delta_{\mathrm{J2016}}$ &  Declination (Dec, degrees) & 11.562632056(3) & \gaia{} EDR3\\
~~~$\mu_{\alpha}$ &  Proper motion (RA, \unit{mas/yr}) &  $23.24\pm0.02$ & \gaia{} EDR3\\
~~~$\mu_{\delta}$ &  Proper motion (Dec, \unit{mas/yr}) & $-38.04\pm0.01$ & \gaia{} EDR3 \\
~~~$d$ &  Distance in pc  & $134.61\pm0.36$ & Bailer-Jones \\
~~~\(A_{V,max}\) & Maximum visual extinction & 0.01 & Green\\
\multicolumn{4}{l}{\hspace{-0.2cm} \textit{Optical and near-infrared magnitudes:}}  \\
~~~$B$ & Johnson B mag & $15.769\pm0.159$ & APASS\\
~~~$V$ & Johnson V mag & $14.395\pm0.056$ & APASS\\
~~~$g^{\prime}$ &  Sloan $g^{\prime}$ mag  & $15.136\pm0.069$ & APASS\\
~~~$r^{\prime}$ &  Sloan $r^{\prime}$ mag  & $13.802\pm0.065$ & APASS \\
~~~$i^{\prime}$ &  Sloan $i^{\prime}$ mag  & $13.068\pm0.074$ & APASS \\
~~~$T$  & \tess{} magnitude & $12.678\pm0.007$  & Stassun \\
~~~$J$ & $J$ mag & $11.466\pm0.023$ & 2MASS\\
~~~$H$ & $H$ mag & $10.749\pm0.024$ & 2MASS\\
~~~$K_s$ & $K_s$ mag & $10.587\pm0.025$ & 2MASS\\
~~~$W1$ &  WISE1 mag & $10.488\pm0.022$ & WISE\\
~~~$W2$ &  WISE2 mag & $10.541\pm0.021$ & WISE\\
~~~$W3$ &  WISE3 mag & $10.436\pm0.089$ & WISE\\
\multicolumn{4}{l}{\hspace{-0.2cm} \textit{Spectroscopic Parameters$^a$:}}\\
~~~$T_{\mathrm{eff}}$ &  Effective temperature in \unit{K} & \stteff{} & This work\\
~~~$\mathrm{[Fe/H]}$ &  Metallicity in dex & $0.38\pm0.04$ & This work\\
~~~$\log(g)$ & Surface gravity in cgs units & $4.67\pm0.12$ & This work\\
\multicolumn{4}{l}{\hspace{-0.2cm} \textit{Model-Dependent Stellar SED and Isochrone fit Parameters$^b$:}}\\
~~~$T_{\mathrm{eff}}$ &  Effective temperature in \unit{K} & $3927\pm{37}$ & This work\\
~~~$\mathrm{[Fe/H]}$ & Metallicity in dex & $0.338^{+0.072}_{-0.066}$ & This work \\
~~~$\log(g)$ &  Surface gravity in cgs units & $4.669_{-0.017}^{+0.018}$ & This work \\
~~~$M_*$ &  Mass in $M_{\odot}$ & $0.639\pm0.023$ & This work\\
~~~$R_*$ &  Radius in $R_{\odot}$ & $0.612_{-0.012}^{+0.013}$ & This work\\
~~~$L_*$ &  Luminosity in $L_{\odot}$ & $0.0803_{-0.0018}^{+0.0019}$ & This work\\
~~~$\rho_*$ &  Density in $\unit{g/cm^{3}}$ & $3.92^{+0.22}_{-0.21}$ & This work\\
~~~Age & Age in Gyrs & $7.1^{+4.4}_{-4.8}$ & This work\\
\multicolumn{4}{l}{\hspace{-0.2cm} \textit{Other Stellar Parameters:}}           \\
~~~$v \sin i_*$ &  Rotational velocity in \unit{km/s}  & $< 2 \kms{}$ & This work\\
~~~$\Delta RV$ &  ``Absolute'' radial velocity in \unit{km/s} & $9.67\pm0.08$ & This work\\
~~~$U, V, W$ &  Galactic velocities (Barycentric) in \unit{km/s} &  $-2.22\pm0.08, -30.20\pm0.11, -1.24\pm0.01$ & This work\\
~~~$U, V, W^c$ &  Galactic velocities (LSR) in \unit{km/s} &  $8.89\pm0.72, -17.96\pm0.48, 6.01\pm0.36$ & This work\\
\enddata
\tablenotetext{}{References are: Stassun \citep{stassun_tess_2018}, 2MASS \citep{cutri_2mass_2003}, \gaia{} EDR3 \citep{gaia_collaboration_gaia_2020}, Bailer-Jones \citep{bailer-jones_estimating_2018}, Green \citep{green_3d_2019}, APASS \citep{henden_apass_2018}, WISE \citep{wright_wide-field_2010}}
\tablenotetext{a}{Derived using the HPF spectral matching algorithm from \cite{stefansson_sub-neptune-sized_2020}}
\tablenotetext{b}{{\tt EXOFASTv2} derived values using MIST isochrones with the \gaia{} parallax and spectroscopic parameters in $a$) as priors.}
\tablenotetext{c}{The barycentric UVW velocities are converted into local standard of rest (LSR) velocities using the constants from \cite{schonrich_local_2010}.}

\end{deluxetable*}

\subsection{Spectroscopic Parameters with HPF-SpecMatch}
Using the method described in \cite{stefansson_sub-neptune-sized_2020}, we use the HPF spectra to estimate the $T_{\mathrm{eff}}$, $\log g$, and [Fe/H] values of the host star. This is based on \texttt{SpecMatch-Emp} algorithm from \cite{yee_precision_2017}, where we compare the high resolution HPF spectra of TOI-532 to a library of high S/N as-observed HPF spectra, which consists of slowly-rotating reference stars with well characterized stellar parameters from \cite{yee_precision_2017}. 

We shift the observed target spectrum to a library wavelength scale and rank all of the targets in the library using a $\chi^2$ goodness-of-fit metric. After this initial $\chi^2$ minimization step, we pick the five best matching reference spectra (in this case: BD+29 2279, GJ 134, GJ 205, HD 28343, HD 88230) to construct a weighted spectrum using their linear combination to better match to the target spectrum (Jones et al. 2021 in prep.). In this step, each of the five stars receives a best-fit weight coefficient. We then assign the target stellar parameter $T_{\mathrm{eff}}$, $\log g$, and \replaced{Fe/H}{[Fe/H]} values as the weighted average of the five best stars using the best-fit weight coefficients. Our final parameters are listed in \autoref{tab:stellarparam}, and are derived from the HPF order spanning 8670 -- 8750 \AA. As an additional check, we performed a similar library comparison using 6 other HPF orders which have low telluric contamination, and obtain consistent stellar parameters across them. Our error estimates are obtained from using the cross-validation method, as described by \cite{stefansson_sub-neptune-sized_2020}.  During both optimization steps, we account for any potential $v \sin i$ broadening by artificially broadening the library spectra with a $v \sin i$ broadening kernel \citep{gray_observation_1992} to match the rotational broadening of the target star. For TOI-532, we did not need significant rotational broadening, and therefore place an upper limit of $v \sin i < 2 \unit{km/s}$, which is the lower limit of measurable $v \sin i$ values given HPF's spectral resolving power of $R \sim 55,000$. 

\subsection{Model-Dependent Stellar Parameters}\label{sec:stellarparams}

In addition to the spectroscopic stellar parameters derived above, we use \texttt{EXOFASTv2} \citep{eastman_exofast_2013} to model the SED of TOI-532  to derive model-dependent constraints on the stellar mass, radius, and age of the star. For the spectral energy distribution (SED) fit, \texttt{EXOFASTV2} uses the BT-NextGen stellar atmospheric models \citep{allard_models_2012}.  We assume Gaussian priors on the (i) 2MASS \(JHK\) magnitudes, (ii) SDSS \(g^\prime r^\prime i^\prime\) and Johnson \(B\) and \(V\) magnitudes from APASS, (iii) \textit{Wide-field Infrared Survey Explorer} magnitudes $W1$, $W2$, and $W3$, \citep[][]{wright_wide-field_2010}, (iv) spectroscopically-derived host star effective temperature, surface gravity, and metallicity, and (v) distance estimate from \cite{bailer-jones_estimating_2021}. We apply a uniform prior on the visual extinction and place an upper limit using estimates of Galactic dust by \cite{green_3d_2019} (Bayestar19) calculated at the distance determined by \cite{bailer-jones_estimating_2021}. We convert the Bayestar19 upper limit to a visual magnitude extinction using the \(R_{v}=3.1\) reddening law from \cite{fitzpatrick_correcting_1999}.

 We use \texttt{GALPY} \citep{bovy_galpy_2015} to calculate the \textit{UVW} velocities in the barycentric frame\footnote{With \textit{U} towards the Galactic center, \textit{V} towards the direction of Galactic spin, and \textit{W} towards the North Galactic Pole \citep{johnson_calculating_1987}.}, which along with the BANYAN tool \citep{gagne_banyan_2018} classify TOI-532 as a field star in the thin disk with very high probability \citep{bensby_exploring_2014}.


\subsection{Estimating Rotation Period}
We note that the \tess{} photometry (PDCSAP undetrended photometry shown in \autoref{fig:tess_lc}) is relatively flat, and shows no flaring activity. We also run a generalized Lomb Scargle (GLS) periodogram \citep{zechmeister_generalised_2009} on the \tess{} photometry using its \texttt{astropy} implementation, and find no significant peaks with a False Alarm Probability $\> 1 \%$\footnote{The PDCSAP photometry from \tess{} flattens variability on timescales longer than about 10 days \citep{jenkins_tess_2016}, and therefore our search using the \tess{} photometry is insensitive to stellar rotation periods longer than this.}. This is consistent with an inactive star with a long rotation period.

\section{Joint Fitting of Photometry and RVs}\label{sec:joint}

\begin{figure}[!t] 
\centering
\includegraphics[width=0.45\textwidth]{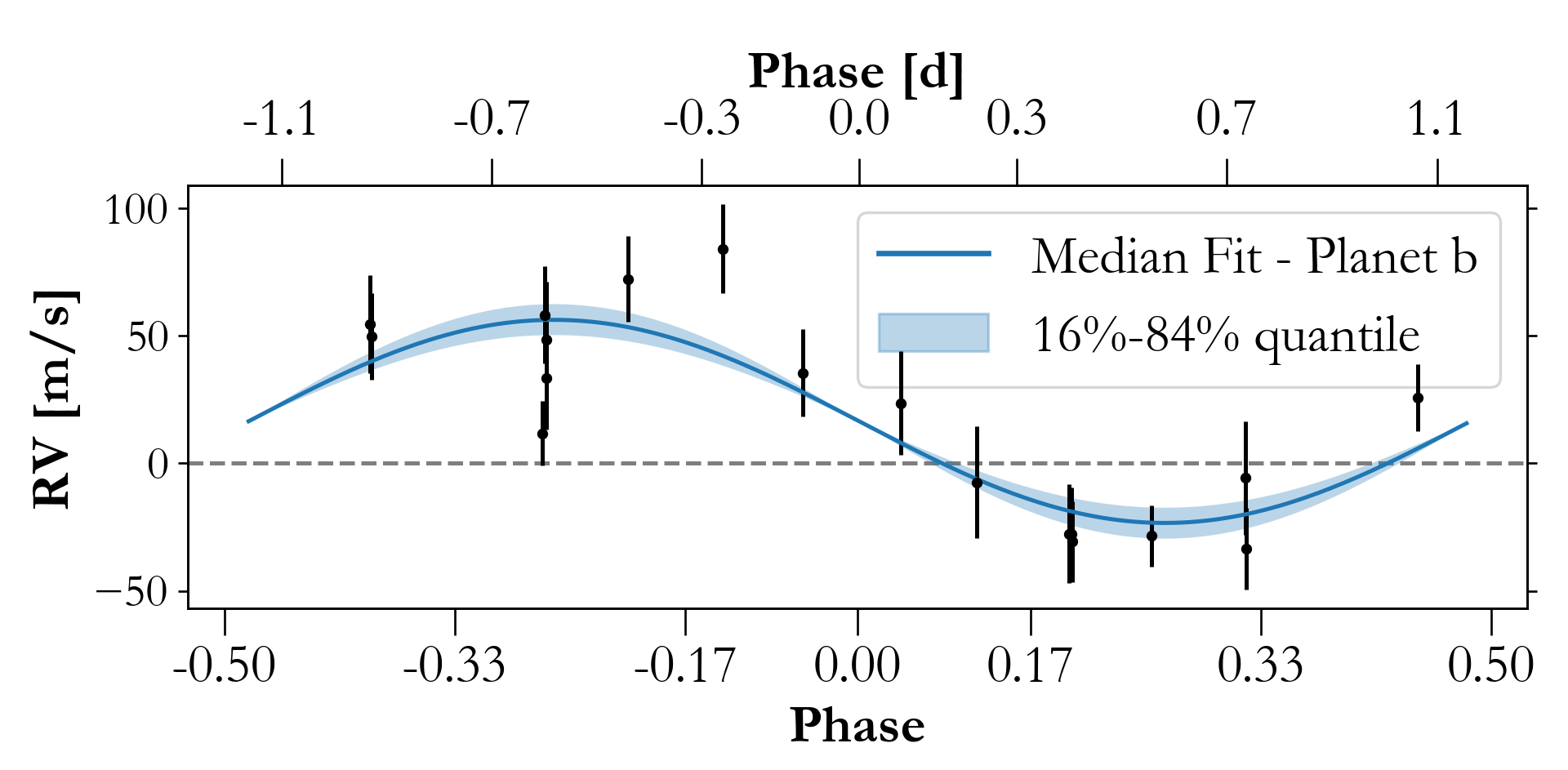}
\caption{HPF RV observations phase folded on the best fit orbital period from the joint fit from Section \ref{sec:joint}. The best fitting model is shown in the solid line, whereas the $1\sigma$ confidence intervals are shown in blue.} \label{fig:RVphase}
\end{figure}

We perform a joint fit of all the photometry (TESS + ground based sources), and the RVs using the \texttt{Python} packge \texttt{exoplanet}, which uses \texttt{PyMC3} the Hamiltonian Monte Carlo (HMC) package \citep{salvatier_probabilistic_2016}.

The \texttt{exoplanet} package uses \texttt{starry} \citep{luger_starry_2019, agol_analytic_2020} to model the planetary transits, using the analytical transit models from \cite{mandel_analytic_2002}, which includes a quadratic limb darkening law. These limb darkening priors are implemented in \texttt{exoplanet} using the reparameterization suggested by \cite{kipping_efficient_2013} for uninformative sampling. We fit each phased transit shown in \autoref{fig:transits} with separate limb darkening coefficients\footnote{We also try fitting the photometry with a single set of limb darkening coefficients for all the transits, and obtain similar results.}. In the photometric model we include a dilution factor for the \tess{} photometry, \(D\), to represent the ratio of the out-of-transit flux of TOI-532 to that of all the stars within the \tess{} aperture, that has not been corrected for.  We assume that the higher spatial resolution ground based photometry has no dilution, since we use the ground based transits to estimate the dilution in the \tess{} photometry. We assume the transit depth is identical in all bandpasses and use our ground-based transits to determine the dilution required in the \tess{} data to be $D_{TESS}$ $= 0.92\pm{0.06}$; including which, gives us a radius of \plradius{} $R_{\oplus}$. If the blending effects due to background stars are correctly accounted for by the SPOC pipeline, we expect this dilution term to be close to 1. 

We model the RVs using a standard Keplerian model.
We try an eccentric joint fit to the photometry and RVs, and obtain an eccentricity consistent with a circular orbit at $\sim 1 \sigma$. Considering the Lucy-Sweeney bias \citep{lucy_spectroscopic_1971}, we adopt a circular orbit by fixing the eccentricity to 0, and the argument of periastron to 90$^{\circ}$. For both the photometry and RV modeling, we include a simple white-noise model in the form of a jitter term that is added in quadrature to the error bars of each data set.

We use \texttt{scipy.optimize} to find the initial maximum a posteriori (MAP) parameter estimates, which are then used as the initial conditions for parameter estimation using  "No U-Turn Sampling" \citep[NUTS,][]{hoffman_no-u-turn_2014}, implemented for the HMC sampler \texttt{PyMC3}, where we check for convergence using the Gelman-Rubin statistic \citep[$\hat{\text{R}} \le 1.1$;][]{ford_improving_2006}. We also run a joint fit using \texttt{juliet} \citep{espinoza_juliet_2019}, and verify that we obtain fit parameters similar to those from \texttt{exoplanet}.

The host stellar density constrained from the transit fit to the \tess{} photometry \citep{seager_unique_2003} is consistent with that obtained from the SED fit for an M0 host star (Section \ref{sec:stellarparams}). The final derived planet parameters are shown in \autoref{tab:planetprop}, and the phased HPF RVs are shown in \autoref{fig:RVphase}. We obtain a mass for TOI-532b of \plmass{} M$_\oplus$, and a radius of \plradius{} R$_\oplus$.

\begin{deluxetable*}{llc}
\tablecaption{Derived Parameters for the TOI-532 System \label{tab:planetprop}}
\tablehead{\colhead{~~~Parameter} &
\colhead{Units} &
\colhead{Value}
}
\startdata
\sidehead{Orbital Parameters:}
~~~Orbital Period\dotfill & $P$ (days) \dotfill & 2.3266508$\pm0.0000030$\\
~~~Eccentricity\dotfill & $e$ \dotfill & 0 (fixed) \\
~~~Argument of Periastron\dotfill & $\omega$ (degrees) \dotfill & 90 (fixed) \\
~~~Semi-amplitude Velocity\dotfill & $K$ (m/s)\dotfill &
39.82$^{+6.15}_{-5.98}$\\
~~~Systemic Velocity$^a$\dotfill & $\gamma$ (m/s)\dotfill & 
16.42$^{+5.04}_{-4.83}$\\
~~~RV trend\dotfill & $dv/dt$ (\unit{m/s/yr})   & 0.35$^{+5.08}_{-4.99}$       \\ 
~~~RV jitter\dotfill & $\sigma_{\mathrm{HPF}}$ (m/s)\dotfill & 11.43$^{+6.62}_{-8.84}$\\
\sidehead{Transit Parameters:}
~~~Transit Midpoint \dotfill & $T_C$ (BJD\textsubscript{TDB})\dotfill & 2458470.576777$^{+0.000860}_{-0.000902}$\\
~~~Scaled Radius\dotfill & $R_{p}/R_{*}$ \dotfill & 
$0.0877\pm0.0016$\\
~~~Scaled Semi-major Axis\dotfill & $a/R_{*}$ \dotfill & 10.49$^{+0.25}_{-0.23}$\\
~~~Orbital Inclination\dotfill & $i$ (degrees)\dotfill & 88.08$^{+0.51}_{-0.41}$\\
~~~Transit Duration\dotfill & $T_{14}$ (days)\dotfill & 0.0728$\pm$0.001\\
~~~Photometric Jitter$^b$ \dotfill & $\sigma_{TESS}$ (ppm)\dotfill & $ 76_{-45}^{+66}$\\
~~~ & $\sigma_{\mathrm{RBO20201207}}$ (ppm)\dotfill & $895_{-584}^{+578}$\\
~~~ & $\sigma_{\mathrm{TMMT20201215}}$ (ppm)\dotfill & $434_{-300}^{+482}$\\
~~~ & $\sigma_{\mathrm{LCRO20210104}}$ (ppm)\dotfill & $540_{-382}^{+697}$\\
~~~ & $\sigma_{\mathrm{TMMT20210104}}$ (ppm)\dotfill & $823_{-585}^{+789}$\\
~~~ & $\sigma_{\mathrm{ARCTIC20210201}}$ (ppm)\dotfill & $770_{-98}^{+101}$\\
~~~Dilution$^c$\dotfill & $D_{\mathrm{TESS}}$ \dotfill & $0.92\pm0.06$\\
\sidehead{Planetary Parameters:}
~~~Mass\dotfill & $M_{p}$ (M$_\oplus$)\dotfill &  \plmass{}\\
~~~Radius\dotfill & $R_{p}$  (R$_\oplus$) \dotfill& \plradius{}\\
~~~Density\dotfill & $\rho_{p}$ (g/\unit{cm^{3}})\dotfill & $1.72\pm0.31$\\
~~~Semi-major Axis\dotfill & $a$ (AU) \dotfill & $0.0296\pm0.00035$\\
~~~Average Incident Flux$^d$\dotfill & $\langle F \rangle$ (\unit{10^5\ W/m^2})\dotfill & 1.28$\pm$0.11 \\
~~~Planetary Insolation& $S$ (S$_\oplus$)\dotfill & $94.1\pm8.0$ \\
~~~Equilibrium Temperature$^e$ \dotfill & $T_{\mathrm{eq}}$ (K)\dotfill & $867\pm18$\\
\enddata
\tablenotetext{a}{In addition to the Absolute RV from \autoref{tab:stellarparam}.}
\tablenotetext{b}{Jitter (per observation) added in quadrature to photometric instrument error.}
\tablenotetext{c}{Dilution due to presence of background stars in \tess{} aperture, not accounted for in the PDCSAP flux.}
\tablenotetext{d}{We use a Solar flux constant = 1360.8 W/m$^2$, to convert insolation to incident flux.}
\tablenotetext{e}{We assume the planet to be a black body with zero albedo and perfect energy redistribution to estimate the equilibrium temperature. }

\normalsize
\end{deluxetable*}

\section{Discussion}\label{sec:discussion}

\begin{figure*}[!t]
\fig{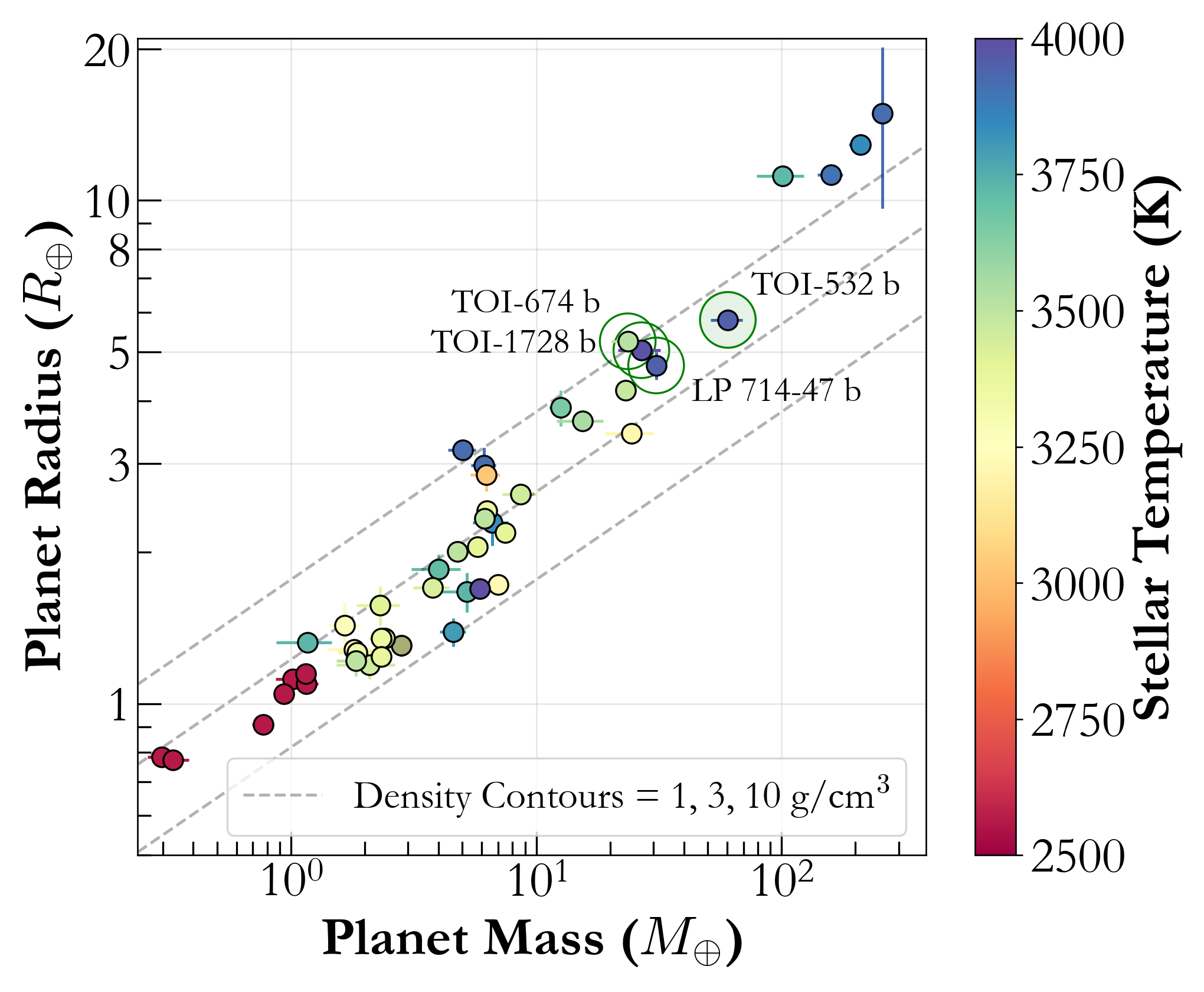}{0.45\textwidth}{{\small a) Mass--Radius plane}}    \label{fig:RadiusMass}
\fig{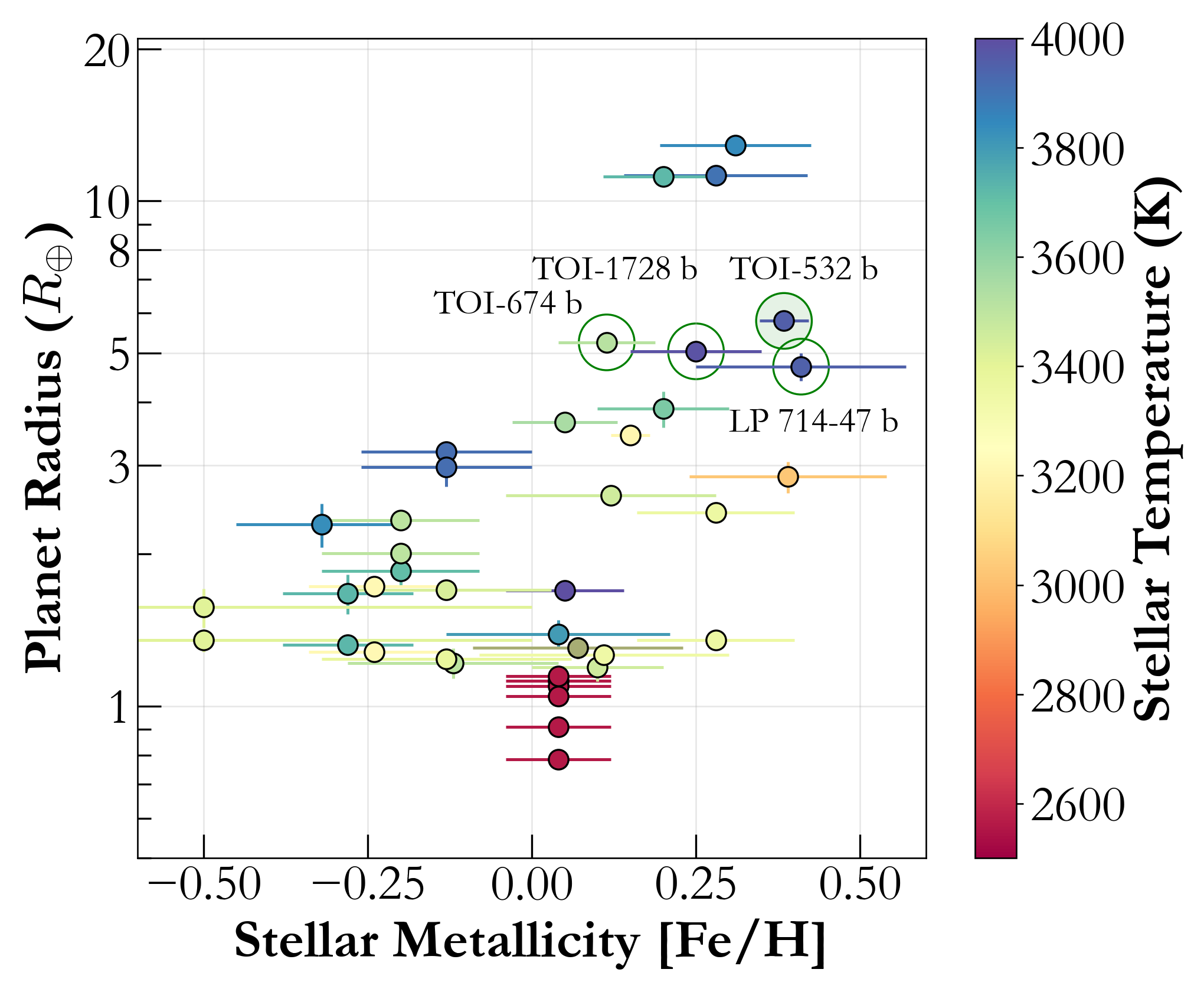}{0.45\textwidth}{ \small b) Metallicity--Radius plane} \label{fig:RadiusMetallicity}
\caption{\small We show TOI-532b (circled) in different planet parameter space along with other M dwarf planets  with mass measurements at $> 3 \sigma$. \textbf{a)}  Mass--Radius plane for M dwarf planets. We include contours of density 1, 3, 10 g/cm$^3$, where the markers are colour coded by \teff{}. \textbf{b)} The metallicity of the host stars for the same planets. We note that all four super Neptunes highlighted in this plot are orbiting metal-rich early type M dwarfs.}\label{fig:PlanetParameters}
\end{figure*}

\begin{figure*}[!t]
\fig{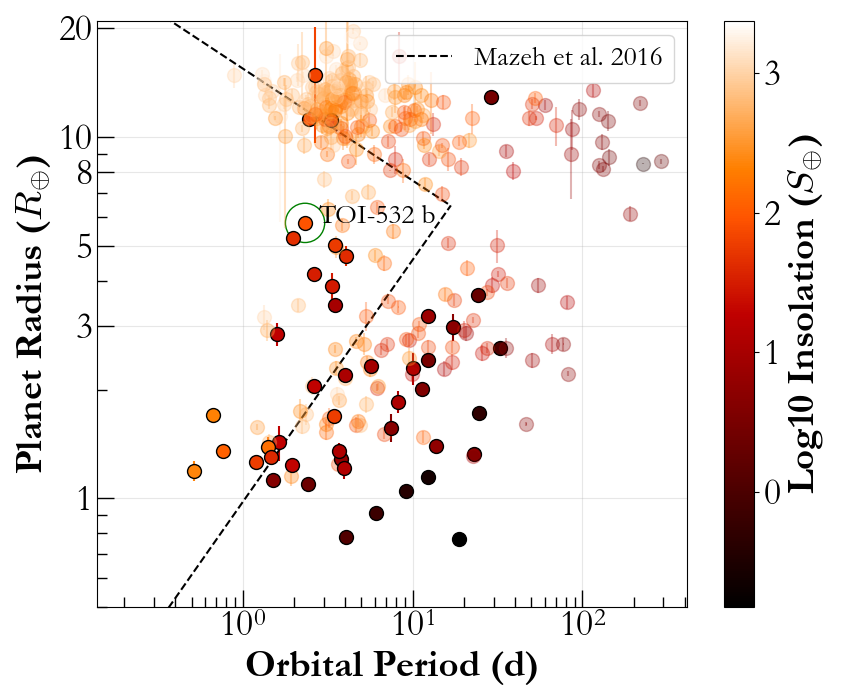}{0.45\textwidth}{{\small a) Radius--Period plane}}    \label{fig:NeptuneDesertPeriod}
\fig{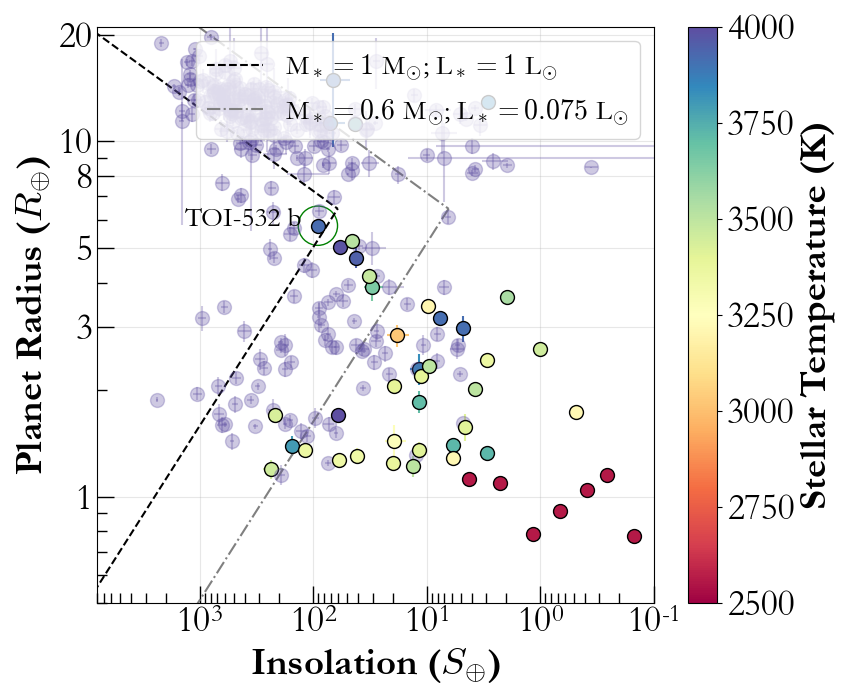}{0.45\textwidth}{ \small b) Radius--Insolation plane} \label{fig:NeptuneDesertInsolation}
\caption{\small We note the location of TOI-532b in the Neptune desert \citep{szabo_short-period_2011} along with a sample of transiting exoplanets that have their masses measured. \textbf{a)} The sample in the Radius--Period plane is colour coded by the log10 insolation, where the M dwarf planets are solid whereas the rest are shown to be translucent. The nominal Neptune desert boundaries from \cite{mazeh_dearth_2016} are denoted with dashed lines. \textbf{b)} We show TOI-532 in the Radius--Insolation plane, where M dwarf planets are coloured according to their \teff{}, with planets orbiting other spectral type host stars denoted in purple. We use the Neptune Desert boundaries from \cite{mazeh_dearth_2016} in Radius--Period plane to calculate similar boundaries in the Insolation plane assuming 1 $M_{\odot}$ and 1 $L_{\odot}$ (Solar type star) in black, and with 0.6 $M_{\odot}$ and 0.075 $L_{\odot}$ (M0 star) in grey for representative purposes. Even though TOI-532 is placed in the middle of the Neptune desert in the Radius--Period plane, we note that in the Radius--Insolation plane it is placed by the edge of the desert, highlighting the importance of considering the insolation fluxes for planetary evolution.}\label{fig:NeptuneDesert}
\end{figure*}

\subsection{Giant Planet Dependence on Host Star Metallicity}
In \autoref{fig:PlanetParameters}a we show TOI-532 b with respect to other M dwarf exoplanets with mass measurements at 3$\sigma$ or higher. The data is taken from the NASA Exoplanet Archive \citep{akeson_nasa_2013}, and includes recent M dwarf transiting planets discovered by TESS. TOI-532b has properties similar to three recent super Neptunes discovered by TESS that orbit M dwarf stars - TOI-1728b \citep{kanodia_toi-1728b_2020}, LP 714-417b \citep[TOI-442 b;][]{dreizler_carmenes_2020}, and TOI-674b \citep{murgas_toi-674b_2021}. TOI-532b represents the largest and most massive Super Neptune found orbiting an M dwarf. 

TOI-532b orbits a metal-rich M star, similar to the other gas giants found around M dwarfs (\autoref{fig:PlanetParameters}b). \replaced{\cite{petigura_california-keplersurvey_2018} suggest that the positive metallicity correlation for gas giants can be explained,}{This positive metallicity correlation favours the core-accretion formation mechanism \citep{pollack_formation_1996, schlaufman_evidence_2018}; which can be explained} if these gas giants formed due to the collisions of 10 $M_{\oplus}$ cores \added{\citep{petigura_california-keplersurvey_2018}}. The probability of formation of these cores increases with metallicity, and therefore it should be easier to form such gaseous planet cores around metal-rich stars, before the protoplanetary disk depletes \citep{ida_toward_2004-1}. In-situ formation of these gas giants at such orbital periods (and hence orbital separations) also requires super-Solar metallicity protoplanetary disks to provide enough material for the formation of their cores \replaced{\citep{dawson_metallicity_2015}}{\citep{dawson_metallicity_2015, boley_situ_2016, batygin_situ_2016}}.

An alternative to the accretion theories of formation \deleted{\citep{pollack_formation_1996}}, is gravitational instability \citep[GI; ][]{boss_giant_1997}. This has been proposed to explain the formation of gas giants around these low mass stars \citep{boss_rapid_2006}, especially the mid-to-late M dwarfs \citep{morales_giant_2019}. The amount of material available in these disks would be too little to form cores that are massive enough to accrete gaseous envelopes from the disk before it gets depleted \citep{laughlin_core_2004}, lending credibility to GI as a potential formation mechanism. The discovery of gas giants such as TOI-532b, adds to the sparse population of these objects around M dwarfs, which can ultimately help differentiate between these two competing theories.

\subsection{Neptune desert}
\autoref{fig:NeptuneDesert} shows the Neptune desert   which is characterized by a dearth of planets. We highlight the location of TOI-532b in the Neptune desert \citep{mazeh_dearth_2016} in the Radius--Period plane (\autoref{fig:NeptuneDesert}a), where it falls in the middle of this region. The figure includes transiting exoplanets with mass measurements, coloured according to their insolation, with the M dwarf planets shown as solid markers, whereas those orbiting other spectral types hosts are translucent. Different processes have been proposed to explain this feature, which include photoevaporation \citep{owen_photoevaporation_2018, ionov_survival_2018}, and high eccentricity migration \citep{matsakos_origin_2016}. 

Although typically parameterized in terms of orbital period, it is important to consider that in a combined sample of FGK and M dwarf host stars, the bolometric insolation can differ by more than an order of magnitude for similar orbital separations (e.g., between a G type host, and an early M dwarf). \cite{mcdonald_sub-neptune_2019} suggest that this variation in the bolometric luminosity is the primary reason for the discrepancies in the location of the Neptune desert as a function of spectral type. Therefore, we also plot TOI-532 in the Radius--Insolation plane (\autoref{fig:NeptuneDesert}b), and include the desert boundaries from \cite{mazeh_dearth_2016} which were estimated using a predominantly FGK planet sample. We include these in the Insolation--Radius plane assuming a Solar mass and luminosity, and also assuming an M0 host star. While TOI-532 is located inside the Neptune desert in the Radius--Period plane, when accounting for the incident insolation, it is located by the edge of this desert. We therefore suggest that in order to compare a sample of planets across spectral types FGK, and M, the Neptune desert should be characterized in the Radius--Insolation plane.

The under-density of planets in this highly irradiated region has often been attributed to atmospheric escape due to photoevaporation \citep{owen_photoevaporation_2018}. The rate and efficacy of photoevaporation is highly dependent on the X-ray and ultraviolet flux (XUV) from the host star; where a planet around a mid-type M dwarf can receive 100x more integrated X-ray flux than a solar type star (for the same insolation). When the frequency distribution of these gaseous planets is considered as a function of lifetime integrated X-ray flux, most of the variability between spectral types is accounted for \citep{mcdonald_sub-neptune_2019}.

Characterization of planets such as TOI-532b, which lie within the Neptune desert, can help provide constraints on the potential formation mechanisms responsible for clearing out the Neptune desert. Estimating the fraction of H/He within its atmosphere would help bound the extent of photoevaporation, and its role in sculpting this desert. TOI-532b helps increase the small sample of planets situated inside this desert. Comparing the stellar (metallicity, age, stellar mass) and planetary parameters (density, planetary mass) for the sample of planets inside the desert to the larger exoplanet sample can help highlight potential formation mechanisms; perhaps as an extension to the radius valley \citep{fulton_california-kepler_2017}, and it's dependence on various stellar properties \citep{owen_metallicity-dependent_2018, cloutier_characterization_2019, berger_gaia-kepler_2020, van_eylen_masses_2021}.

\subsection{Planetary Composition and Photoevaporation}
We use the giant planet models from \cite{fortney_planetary_2007} to estimate a core mass of $\sim 36$ \earthmass{} for TOI-532b, corresponding to an atmospheric mass (H/He) fraction of $\sim 25 \%$.
Super Neptunes such as TOI-532b present an intermediate population of gaseous planets between sub-Neptunes \citep{bean_nature_2021} and Jovian planets \citep{mordasini_imprint_2016, dawson_origins_2018}. A subset of these Super Neptunes with equilibrium temperatures between 800--1200 K span the range where models predict a transition from methane dominated atmospheres to carbon monoxide \citep{guzman-mesa_information_2020}. Characterizing the atmospheres of planets such as TOI-532b with equilibrium temperatures of $\sim 850$ K by constraining their C/H and C/O ratios, can help place constraints on their formation history as well as atmospheric chemistry \citep{madhusudhan_atmospheric_2017}.

TOI-532 is relatively faint (J = 11.46), but is still accessible from 10-m class telescopes \citep{tamura_infrared_2012, kotani_infrared_2018}, as a potential target for detecting atmospheric escape using the He 10830 \AA~ triplet. Considering the small number of suitable targets for such a measurement, we discuss the possibility of detecting atmospheric escape from TOI-532b.  It is useful to compare TOI-532b to a similar planet with such a detection---GJ 3470b \citep{ninan_evidence_2020,palle_he_2020}---and also to a planet without a He 10830 \AA~ detection, TOI-1728b \citep{kanodia_toi-1728b_2020}. In the energy limited mass outflow regime\footnote{The energy limited regime is a reasonable assumption here since the gravitational potential of this planet is 12.81 erg g$^{-1}$ \citep[log10 (GM/R);][]{salz_energy-limited_2016} . This is not a system with a low density upper-atmosphere, like those seen in planets with higher gravitational potential ($>$13.3 erg g$^{-1}$)}, the exosphere outflow is proportional to the irradiated extreme ultra violet (EUV) flux and inversely proportional to the planet density. TOI-532 is an earlier M0 star than the M1.5 GJ 3470, with its spectral type more favourable with higher EUV radiation. However, TOI-532 is an older (and quieter) $7.2^{+4.6}_{-4.7}$ Gyr star, while GJ 3470 is relatively young at $\sim3$ Gyr\footnote{GJ 3470b stellar and planetary parameters are from \cite{kosiarek_bright_2019}.}. If we consider the EUV flux from the host star to be similar, due to the larger radius of the host star, the EUV irradiance on TOI-532b is 1.6 times that of GJ 3470b, which can make up for the 1.8 times higher density of TOI-532b than GJ 3470b. Thus, \textit{if} the EUV flux of TOI-532 (7 Gyr, M0) is similar to GJ 3470 (3 Gyr, M1.5), we could expect a similar exosphere evaporation and mass outflow in TOI-532b like in GJ 3470b. Under this condition, He 10830 \AA~ absorption during transit is a good probe to detect any signatures of outflow from TOI-532.  

Conversely, the other planet TOI-1728b has a host star very similar to TOI-532 in both spectral type and age. TOI-532 orbits 1.25 times closer than TOI-1728b, and it is 1.5 times denser than TOI-1728b. Therefore, from a simple scaling relationship we expect the mass outflows in them to be only slightly less or very similar. That being said, TOI-1728b had a null detection of He 10830 \AA~ with an upper limit of 1.1\% \citep{kanodia_toi-1728b_2020}. We therefore note that though the planetary parameters are amenable, the plausibility of a detectable outflow from this super Neptune hinges on the EUV irradiation environment of the host star.


\section{Summary}\label{sec:summary}
In this work, we report the discovery and confirmation of a super Neptune, TOI-532b, orbiting an M0 star in a $\sim$ 2.3 day circular orbit.  We detail the \tess{} photometry, ground-based follow-up photometry, high contrast imaging, and also the RV observations performed using HPF. Furthermore, we discuss how the planet is situated at the edge of the Neptune desert in the Radius--Insolation plane, and discuss potential for He 10830 \AA~ absorption detection using transmission spectroscopy. We also discuss the metallicity correlation for gas giants occurrence, and how it continues down to the M dwarf spectral type.

The discovery and mass measurement of gas giants such as TOI-532b adds to the small sample of such planets around M dwarf host stars, and can help inform theories of planetary formation and evolution. Therefore we encourage future observations to place limits on atmospheric escape using the He 10830 \AA~ transition.

\section{Acknowledgements}
This research made use of Lightkurve, a Python package for Kepler and TESS data analysis (Lightkurve Collaboration, 2018).

This paper is based on observations obtained from the Las Campanas Remote Observatory that is a partnership between Carnegie Observatories, The Astro-Physics Corporation, Howard Hedlund, Michael Long, Dave Jurasevich, and SSC Observatories.

This work has made use of data from the European Space Agency (ESA) mission
{\it Gaia} (\url{https://www.cosmos.esa.int/gaia}), processed by the {\it Gaia}
Data Processing and Analysis Consortium (DPAC,
\url{https://www.cosmos.esa.int/web/gaia/dpac/consortium}). Funding for the DPAC
has been provided by national institutions, in particular the institutions
participating in the {\it Gaia} Multilateral Agreement.

The Center for Exoplanets and Habitable Worlds is supported by the Pennsylvania State University, the Eberly College of Science, and the Pennsylvania Space Grant Consortium.
These results are based on observations obtained with the Habitable-zone Planet Finder Spectrograph on the HET. We acknowledge support from NSF grants AST 1006676, AST 1126413, AST 1310875, AST 1310885, and the NASA Astrobiology Institute (NNA09DA76A) in our pursuit of precision radial velocities in the NIR. We acknowledge support from the Heising-Simons Foundation via grant 2017-0494.  The Hobby-Eberly Telescope is a joint project of the University of Texas at Austin, the Pennsylvania State University, Ludwig-Maximilians-Universität München, and Georg-August Universität Gottingen. The HET is named in honor of its principal benefactors, William P. Hobby and Robert E. Eberly. The HET collaboration acknowledges the support and resources from the Texas Advanced Computing Center. We thank the Resident astronomers and Telescope Operators at the HET for the skillful execution of our observations with HPF.

We acknowledge support from NSF grants AST-1909506 and AST-1907622 and the Research Corporation for precision photometric observations with diffuser-assisted photometry.

This work was performed under the following financial assistance award 70NANB18H006 from U.S. Department of Commerce, National Institute of Standards and Technology  

This research has made use of the NASA Exoplanet Archive, which is operated by the California Institute of Technology, under contract with the National Aeronautics and Space Administration under the Exoplanet Exploration Program. 
This work includes data collected by the \tess{} mission, which are publicly available from MAST. Funding for the \tess{} mission is provided by the NASA Science Mission directorate. 
Some of the data presented in this paper were obtained from MAST. Support for MAST for non-HST data is provided by the NASA Office of Space Science via grant NNX09AF08G and by other grants and contracts.

This research has made use of the SIMBAD database, operated at CDS, Strasbourg, France, 
and NASA's Astrophysics Data System Bibliographic Services.

Some of the observations in this paper made use of the NN-EXPLORE Exoplanet and Stellar Speckle Imager (NESSI). NESSI was funded by the NASA Exoplanet Exploration Program and the NASA Ames Research Center. NESSI was built at the Ames Research Center by Steve B. Howell, Nic Scott, Elliott P. Horch, and Emmett Quigley.

Part of this research was carried out at the Jet Propulsion Laboratory, California Institute of Technology, under a contract with the National Aeronautics and Space Administration (NASA).

Computations for this research were performed on the Pennsylvania State University’s Institute for Computational and Data Sciences Advanced CyberInfrastructure (ICDS-ACI), including the CyberLAMP cluster supported by NSF grant MRI-1626251.
This work includes data from 2MASS, which is a joint project of the University of Massachusetts and IPAC at Caltech funded by NASA and the NSF.  CIC acknowledges support by NASA Headquarters under the NASA Earth and Space Science Fellowship Program through grant 80NSSC18K1114.  SK would like to acknowledge Monae and Theodora for help with this project.

This research made use of \textsf{exoplanet} \citep{foreman-mackey_exoplanet-devexoplanet_2021} and its
dependencies \citep{agol_analytic_2020, kumar_arviz_2019, robitaille_astropy_2013, astropy_collaboration_astropy_2018, kipping_efficient_2013, luger_starry_2019, the_theano_development_team_theano_2016, salvatier_probabilistic_2016, foreman-mackey_exoplanet_2021}

\facilities{\gaia{}, HET (HPF), \tess{}, TMMT, LCRO, RBO, APO (ARCTIC), WIYN (NESSI), Shane (ShARCS), Exoplanet Archive}
\software{
\texttt{ArviZ} \citep{kumar_arviz_2019}, 
AstroImageJ \citep{collins_astroimagej_2017}, 
\texttt{astroquery} \citep{ginsburg_astroquery_2019}, 
\texttt{astropy} \citep{robitaille_astropy_2013, astropy_collaboration_astropy_2018},
\texttt{barycorrpy} \citep{kanodia_python_2018}, 
\texttt{HxRGproc} \citep{ninan_habitable-zone_2018},
\texttt{ipython} \citep{perez_ipython_2007},
\texttt{juliet} \citep{espinoza_juliet_2019},
 \texttt{lightkurve} \citep{lightkurve_collaboration_lightkurve_2018},
\texttt{matplotlib} \citep{hunter_matplotlib_2007},
\texttt{MRExo} \citep{kanodia_mass-radius_2019},
\texttt{numpy} \citep{oliphant_numpy_2006},
\texttt{pandas} \citep{mckinney_data_2010},
\texttt{PyMC3}\citep{salvatier_probabilistic_2016},
\texttt{scipy} \citep{oliphant_python_2007, virtanen_scipy_2020},
\texttt{SERVAL} \citep{zechmeister_spectrum_2018},
\texttt{starry} \citep{luger_starry_2019, agol_analytic_2020},
\texttt{Theano} \citep{the_theano_development_team_theano_2016}.
}

\bibliography{references}



\end{document}